\pdfoutput=1 
\documentclass[12pt]{article}
\usepackage[top=3cm, left=3cm, right=3cm, bottom=3cm]{geometry}
\usepackage{floatrow}
\usepackage[utf8]{inputenc}
\usepackage[english]{babel}
\usepackage{graphicx}
\usepackage[T1]{fontenc}
\usepackage{amsmath, amssymb}
\usepackage{natbib}
\usepackage{float}
\floatstyle{plaintop}
\restylefloat{table}
\usepackage{multirow}
\usepackage{parskip}
\usepackage{times}
\usepackage[font=small,labelfont=bf]{caption}
\usepackage{subcaption}
 \usepackage[bottom]{footmisc}
\usepackage{hyperref}
\usepackage{multicol}
\usepackage{lipsum}
\usepackage{authblk}
\usepackage{dcolumn}
\usepackage{fontawesome}
\usepackage{animate}
\usepackage{multimedia}
\usepackage{media9}
\usepackage[Export]{adjustbox}
\usepackage{tabularx}
\usepackage{cellspace}

\DeclareCaptionLabelSeparator{pipe}{ $\textbf{|}$ }

\let\oldnl\nl
\newcommand{\nonl}{\renewcommand{\nl}{\let\nl\oldnl}}

\usepackage[title]{appendix}
\usepackage{algorithm2e}

\makeatletter
\renewcommand{\@algocf@capt@plain}{above}
\makeatother

\usepackage{algcompatible}

\SetKwProg{Fn}{Function}{}{end}\SetKwFunction{FRecurs}{FnRecursive}%
\SetAlgoLongEnd

\title{Modeling the Age Pattern of Fertility: An Individual-Level Approach} 

\author[1,2]{Daniel Ciganda \faEnvelopeO }
\author[3]{Nicolas Todd}

\affil[1]{\small{Max Planck Institute for Demographic Research}}
\affil[2]{\small{Statistics Institute, University of the Republic, Montevideo, Uruguay}}
\affil[3]{\small{UMR7206 Eco Anthropologie, Musée de l'Homme - CNRS, Paris, France}}

\date{}

\begin{document}
\setlength{\parindent}{0pt}

\maketitle              
\begin{abstract}

Macro-level modeling is still the dominant approach in many demographic applications because of its simplicity. Individual-level models, on the other hand, provide a more comprehensive understanding of observed patterns; however, their estimation using real data has remained a challenge. The approach we introduce in this article attempts to overcome this limitation. Using likelihood-free inference techniques, we show that it is possible to estimate the parameters of a simple but demographically interpretable individual-level model of the reproductive process from a set of aggregate fertility rates. By estimating individual-level quantities from widely available aggregate data, this approach can contribute to a better understanding of reproductive behavior and its driving mechanisms. It also allows for a more direct link between individual-level and population-level processes. We illustrate our approach using data from three natural fertility populations.

\end{abstract}

\vfill

\section{Introduction}
\label{intro}

Modeling the age distribution of fertility rates is an essential step in a number of demographic applications. When only a tight fit to an observed schedule is required, as in the generation of single-age rates from grouped data, nonparametric models, typically based on splines, tend to produce the best results. No particular price is then placed on whether each model coefficient may be interpreted in any meaningful way. Other applications require models that can fit the data well while also providing a well-defined, ideally small set of parameters that represent quantities that can be interpreted in demographic terms. This typically involves ages summarizing relevant characteristics of the fitted curve, such as the age at which the curve rises above 0, the age at which it peaks, or the age at which it drops down to zero.

The interpretability of parameters is particularly relevant in a forecasting context, where it may prove especially useful to associate a change in the value of the parameters with an underlying behavioral process, such as fertility postponement or the diffusion of contraceptive methods. Forecasting each parameter's change in time is then possible, and potentially yields improved forecasting of the aggregate quantities of interest.

A difficulty associated with most existing parametric models is that the interpretation of their parameters can be elusive \citep{hoem1981experiments}. Even in the best cases, the relationship between mechanisms and parameters is ambiguous and indirect. In essence, macro-level models always operate one level above the behaviors that drive fertility.

A natural solution to this problem is to model at the individual level. This approach involves three steps: 1) develop a model of the reproductive process; 2) use this model to generate synthetic data and compute a set of simulated age-specific fertility rates (ASFR); 3) estimate the parameters of the model by minimizing the distance between simulated and observed rates.   

Although the building blocks have been available for some time, this approach has not yet been systematically explored. 

The basic tools and ideas required for step 1 have long been developed by some of the pioneers of mathematical demography \citep[][notably]{gini1924premieres, Henry1953, sheps1973mathematical}. 

These ideas were later explored in more realistic settings using simulation methods. The objective was then the incorporation of  basic characteristics of the reproductive process like age-dependent fecundability, or to represent change over time in fertility rates \citep[see:][]{ridley1966analytic,Barrett1971, le1993simulation}. The use of simulation methods allowed researchers to complete step 2, by generating synthetic age-specific fertility rates from individual-level models, drawing a direct connection between behavior and aggregate quantities.  

These contributions, however, remained largely theoretical. Parameter values where borrowed from previous studies, or calibrated through trial and error until the simulated data would fit a given target distribution. In other words, estimation procedures remained too fuzzy, if not arbitrary, to decide whether individual-level models could properly be used on observed data.

Interest in these type of models dwindled after the 1980's, just at the same time the first statistical methods that would enable inference on complex simulation models were starting to emerge in other disciplines \cite{rubin1984bayesianly}.

Likelihood-free inference methods, such as Approximate Bayesian Computation (ABC), have been extensively researched since then, providing a robust statistical framework to estimate the parameters of computational models \cite{Beaumont2019}. What we attempt to show in the remainder of this paper is that these advances in statistical computing, that have proved so useful in fields such as ecology and population genetics, also provide the missing piece to the program outlined above. Specifically, these advances allow to infer characteristics of  individual level processes from aggregate fertility rates. More generally, the approach we advocate here provides a clear roadmap to fully integrate behavioral mechanisms in the modeling and forecasting of fertility trends.  

\section{Model}

The simplest form of the reproductive process is the one in which the outcomes of the process (exact timing and number of births) are not a function of individual preferences. Louis Henry called this ``natural fertility'', in an attempt to characterize a setting in which the reproductive process is exclusively determined by physiological factors. 


A model of the reproductive process in a natural fertility setting requires three basic inputs: The moment when the process starts, typically defined as the age at marriage; the risk or probability of a conception, known as fecundability; and the length of the period in which women are not able to conceive following childbirth, know as post-partum amenorrhea.

We represent the reproductive experience of a cohort of women from birth to age 50 using a discrete-time simulation model, in which time advances at fixed increments of one month. The age (in months) at marriage for a woman $i$ is simulated from a lognormal distribution with mean $\mu_{m}$ and standard deviation $\sigma_{m}$.

Every month, married women who are neither pregnant nor amenorrheic are exposed to the risk of a conception, which depends on the woman's age. Whether woman $i$ conceives in a given month when aged $x^*$ is the outcome of a Bernoulli trial with probability $\phi(x^*)$.

We model the age-pattern of fecundability using two polynomial bases:

\begin{equation}
\phi(x) = \phi_{1} \cdot 3x_s(1-x_{s})^2 + \phi_2 \cdot 3x_s^2(1- x_s)
\end{equation}

Where $\phi(x)$ is fecundability at age $x$ in the reproductive window (assumed 10 to 50 years), and $x_s = \frac{x-10}{50-10}$, the age scaled to belong to $[0, 1]$. These basis functions are two of the four Bernstein polynomials of degree 3. The other two [$x_s\mapsto (1-x_s)^3$ and $x_s \mapsto x_s^3$] have no use when representing fecundability, which is 0 at ages 10 and 50. 

\begin{figure}[H]
\centering
\includegraphics[width=0.7\textwidth,height=0.7\textheight,keepaspectratio]{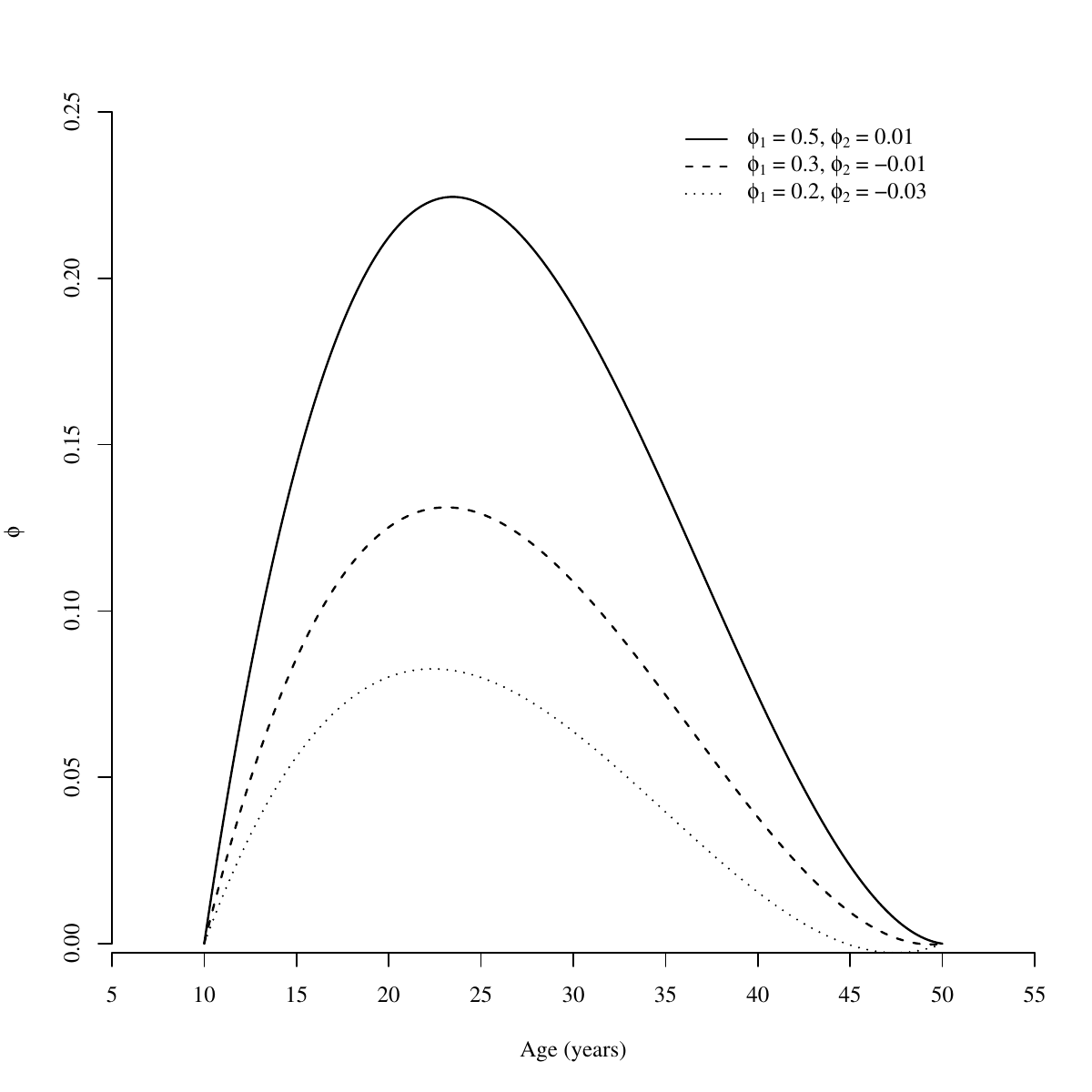}
\caption[Age Distribution of Fecundability for Different Combinations of Parameter Values]%
{\textbf{Age Distribution of Fecundability for Different Combinations of Parameter Values}. Although it depends on only two parameters, the model used to represent the evolution of the risk of conception across the reproductive life course captures a wide range of patterns.}
\label{phi}
\end{figure}

Different combinations of $\phi_1$ and $\phi_2$ generate different age profiles for the risk of conception, as seen in Figure~\ref{phi}. $\phi_1$ has a stronger influence on the level at which fecundability peaks, while $\phi_2$ largely controls the pace at which the risk of conception decays, until permanent sterility is reached. While the specific values of $\phi_1$ and $\phi_2$ are not informative on their own, the resulting age profile of fecundability produce by a given combination reflects both the behavioral (such as the frequency of intercourse) and biological mechanisms that influence the risk of conception.   

\captionsetup[figure]{labelsep=pipe, name={Fig.}}

A conception brings about a state of non-susceptibility (no risk of conception) equal to the duration of the pregnancy, 9 months, plus the duration of post-partum amenhorrea, $\delta$, that must be estimated together with the other parameters of the model.

The model outputs individual reproductive trajectories, from which a set of simulated age-specific fertility rates ${}_{s}f(x)$ can be computed. We then leverage the distance between simulated and observed rates, ${}_{o}f(x)$, to estimate the parameters of our model, namely $(\mu_m, \sigma_m, \phi_1, \phi_2, \delta)$. This procedure is explained in more detail in Section~\ref{estim}, after we describe the data used in our analysis in the next section.

\section{Data}
\label{data}

To illustrate our approach, we use data from three high quality natural fertility populations: Hutterites communities from 1860 to 1914, the French-Canadian population from 1700 - 1750, and a subsample of the Louis Henry Enquête on the French population in the XVIIIth century. All three provide full maternity histories as well as information on the dates of other relevant events in a woman's lifecourse like marriage, death, or death of spouse.

\subsection{Hutterites}

The Hutterites are a Anabaptist community, originated in the XIVth century in the Tyrolean Alps. After a long history of migration in Europe and Asia, they relocated to North America at the end of the XIXth century where they still live today in self-sustained, largely isolated colonies. 

Like many religious communities, the Hutterites oppose the use of birth control methods \citep{Lee1967, Ingoldsby1988}. What made them stand out for demographic analysis was how strictly they adhered to these beliefs to at least the second half of the 20th century. The marital fertility of the Hutterite cohorts born until the early 1900s remained close to the theoretical maximum, with an average of around 10 children per woman, providing researchers with an exceptional opportunity to study the reproductive process under natural fertility conditions \citep{Eaton1953}. 

Another reason why the Hutterites became the gold standard for natural fertility research was their custom of keeping detailed family records. These records, personally checked for consistency by colony preachers, were made available for various scientific studies in the 1950s and 60s \citep{Eaton1953, Mange1964}. These data were extended in the course of one of these studies through follow-up interviews, which resulted in complete maternity histories for 562 families \cite{Sheps1965}. 

\subsection{French-Canadian}

BALSAC is a longitudinal population database that contains information on individuals and families who have lived in Quebec from the 17th century to the present. The information used in this article, that concerns cohorts born before 1750, was gathered using family reconstitution methods from parish registers \citep{Vezina2020}. Given the high quality of these records, the information on XVII Century Quebec populations has become an important reference in the literature on natural fertility \citep[see:][among others]{Clark2020, Larsen2000, Eijkemans2014}.     


\subsection{Henry}

Another important dataset for natural fertility research is the Enquête Louis Henry. This dataset was created by Louis Henry and his coworkers as an effort to draw a nationally representative sample of XVIIIth century France population. It contains data on life events for a sample of 378 parishes for the period 1670 to 1829 from parish registers and administrative records -- the 'anonymous sample'. The families in a sub-sample of 40 rural parishes -- the 'nominative sample' -- have been fully reconstituted  \citep[for a detailed description of the Enquête, see]{seguy2001}.

\subsection{Sample Sizes}

To keep things simple, we model a process without death or union dissolution. Therefore, we restrict our data to ``intact'' marriages, i.e., marriages that do not dissolve by death or separation before the woman reaches age 50. Table~\ref{tab:data} contains the remaining sample sizes after we exclude censored trajectories and records with incomplete information. Records with partial information were recovered in some cases by imputing the month or day of events. The inclusion of imputed data does not significantly affects our results.   

While our simulated data belongs to a single birth cohort, in order to get reliable estimates our empirical data is obtained from multiple cohorts. To minimize potential biases we kept a range of birth cohorts that are as homogeneous as possible with respect to the age at first marriage and the total fertility rate. The cohorts included for each population are also displayed in Table~\ref{tab:data}  

\begin{table}[h!]
\centering
\begin{tabular}{|l|c|c|c|}
\hline
 & Nr. Marriages & Nr. Births & Cohorts \\
\hline
Hutterites & 161 & 1726 & 1860 - 1914 \\
XVIII Century Quebec & 14303 & 110772 & 1722 - 1730 \\
XVII - XVIII Century France & 3235 & 18623 & 1680 - 1760 \\
\hline
\end{tabular}
\caption{Sample Information for the Three Natural Fertility Datasets}
\label{tab:data}
\end{table}

\begin{figure}[H]
\centering
\includegraphics[width=0.7\textwidth,height=0.7\textheight,keepaspectratio]{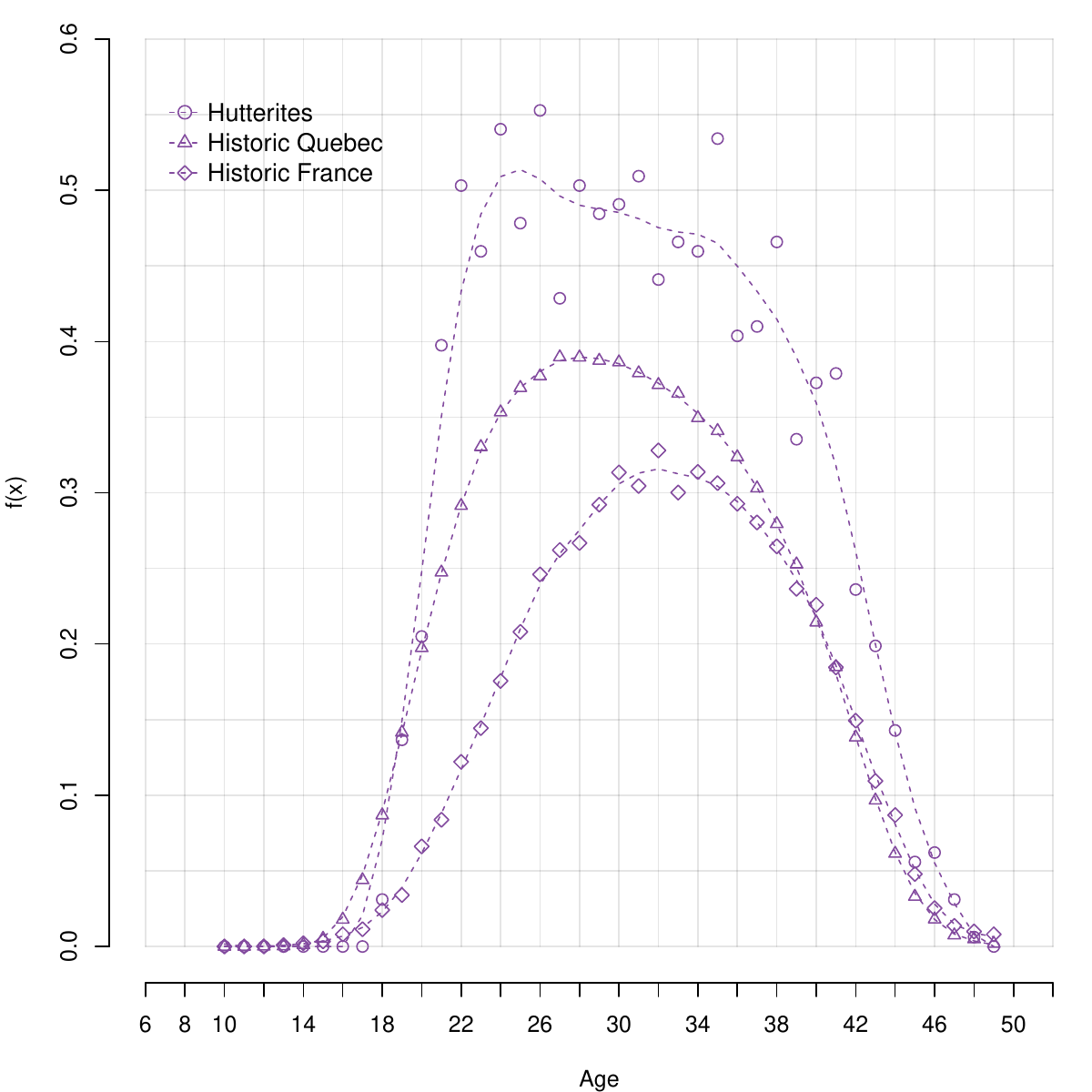}
\caption[Age Specific Fertility Rates for Three Natural Fertility Populations]%
{\textbf{Age Specific Fertility Rates for Three Natural Fertility Populations}. }
\label{obs_fx}
\end{figure}

Figure~\ref{obs_fx} shows the observed age-specific fertility rates for our three natural fertility populations. As expected, the Hutterites show consistently high fertility from age 20, with an average number of children per woman in these cohorts of 10.76. The Hutterite data is also unique in terms of its pattern, with a visible hump around ages 24-26, followed by a decline that is slow until age 35 and accelerates afterwards. The French-Canadian cohorts have the second largest fertility in the three populations, with an average of 7.7 children per woman, followed by the French cohorts with an average of 5.75 children per woman. The shape of the ASFR is more conventional in these last two populations, characterized by a rounded top, which in the case of the French-Canadian cohorts is found around ages 28 and in the case of the French cohorts around age 32. This is not surprising considering the mean ages at marriages for these cohorts which is 20.4 for the Hutterites, 21.1 for the French-Canadian, and 24.1 for the French cohorts       

\captionsetup[figure]{labelsep=pipe, name={Fig.}}

\subsection{Estimation}
\label{estim}

As previously mentioned, our primary goal is to demonstrate that the lack of individual-level data does not have to be an impediment to understanding the behaviors and mechanisms underlying fertility processes. 

When only aggregate data is available, the probability of individual observations cannot be calculated, and model parameters cannot be estimated using a maximum likelihood approach. Instead, we rely on Approximate Bayesian Computation (ABC), a popular likelihood-free approach for the estimation of computational models. 

The objective of ABC is to get a sample approximately drawn from the posterior distribution without ever explicitly calculating the likelihood function. Instead, ABC algorithms simulate data from the model for a set of parameter values, and keep combinations of parameter values that yield simulated data that is close to the observed data, according to a predefined distance metric. 

In the basic ABC rejection algorithm we use, parameter samples are drawn from the prior distribution \citep{tavare1997inferring, pritchard1999population}. While this method is straightforward to implement, it requires a large number of simulations to obtain a reasonable sample from the approximate posterior.

To improve the efficiency of the approach, the parameter values accepted in the first step can be modeled as a function of the simulated data to obtain a better approximation of the posterior distribution, a procedure known in the ABC literature as ``regression adjustment'' \citep{blum2018regression}. For this second step, we use a random forest (RF), as described in the pseudocode below.

\begin{algorithm}
\DontPrintSemicolon

\TitleOfAlgo{ABC + Regression Adjustment}

\textbf{Input:} Prior distribution \(\pi\), model $\mathcal{M}$ for simulation, observed data \(y_{\text{obs}}\), distance threshold \(\epsilon\). \;
\textbf{Output:} Adjusted samples from the posterior distribution \(\{\tilde{\theta}_1, \tilde{\theta}_2, \ldots, \tilde{\theta}_N\}\). \;

\Begin{
    Draw parameter samples, \(\theta^*\), from \(\pi\) \;
    For each \(\theta^*\), simulate data, \(y^*\), from $\mathcal{M}(\theta^*)$ \;
    Compute distance between \(y^*\) and \(y_{\text{obs}}\) \;
    Accept \(\theta^*\) if the distance is less than \(\epsilon\) \;

    With accepted parameters \(\{\theta_1, \theta_2, \ldots, \theta_N\}\) and corresponding simulated data \(\{y_1, y_2, \ldots, y_N\}\): \;
    \Begin{
        Train a Random Forest $m$ that maps $y$ to $\theta$ \;
        
        Adjust the values of the accepted parameters: \;
        \[
        \tilde{\theta}_i  = m(y_{\text{obs}}) + (\theta_i - m(y_i))
        \]
    }
    Return \(\{\tilde{\theta}_1, \tilde{\theta}_2, \ldots, \tilde{\theta}_N\}\) as samples from the adjusted posterior distribution. \;
}
\end{algorithm}

\subsection{Sample Size}

The comparison between simulated and observed data in ABC typically relies on summary statistics, as using the raw data directly is usually impractical or impossible. 

We use the set of age-specific fertility rates for our three natural fertility populations as our target summary statistics. As shown in Table~\ref{tab:data}, the number of individual trajectories used to compute these rates varies widely between the three populations. To adequately represent the uncertainty associated with each sample size, we estimate each population separately, matching the number of marriages in our observed data with the number of marriages used when simulating from the model.

\subsection{Prior choice}

Given that one of our primary goals is to determine whether or not it is possible to infer the parameters of our individual model from information contained on a set of age-specific fertility rates, we use non informative priors for our model's parameters. More specifically, the results presented in the following sections are obtained using uniform priors in a range containing reasonably implausible values for the five parameters in our model.

\section{Results}
\label{results}

\subsection{Model Validation}

We use cross-validation to assess whether the parameters in our model are correctly identified from a set of age-specific fertility rates. This procedure makes use of the set of parameter values $\theta^*$ and their corresponding simulated rates $y^*$. From this set we randomly select the $i^{th}$ simulation as our validation set. The simulated rates in the validation set are taken as pseudo-observed values. Taking into account all simulations except the validation set, the model parameters are then estimated using the procedure described in the previous section. This process is repeated for a subset of 100 randomly chosen simulations and the prediction error is computed as:

\[
\varepsilon_{\text{p}} = \sum_i \left( \frac{(\theta^{\ast}_{i} - \theta_i)^2}{\text{Var}(\theta_i)} \right)
 \]

\begin{figure} [H]
\caption[Results of Model Cross-Validation.]{\textbf{Results of Model Cross-Validation.} The location of each of the red dots displayed in the figure is defined by the mean of the estimated posterior distribution in the $y$ axis and the value of the parameter that generated the simulated results used as pseudo-observed data during the validation process.}
        \begin{subfigure}[t]{0.33\textwidth}
                \centering
                \includegraphics[width=\textwidth]{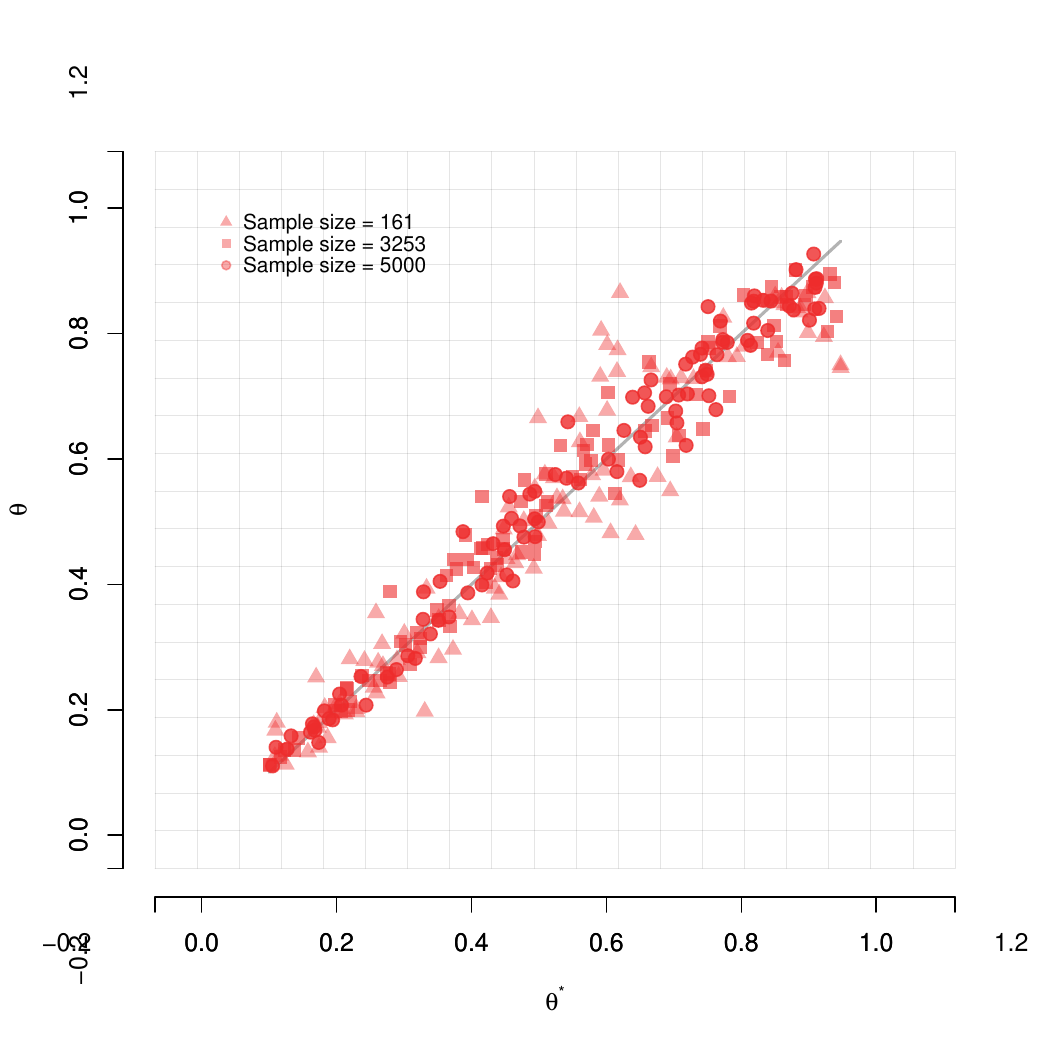}
                \caption{$\phi_{1}$}
        \end{subfigure}%
        \begin{subfigure}[t]{0.33\textwidth}
                \centering
                \includegraphics[width=\textwidth]{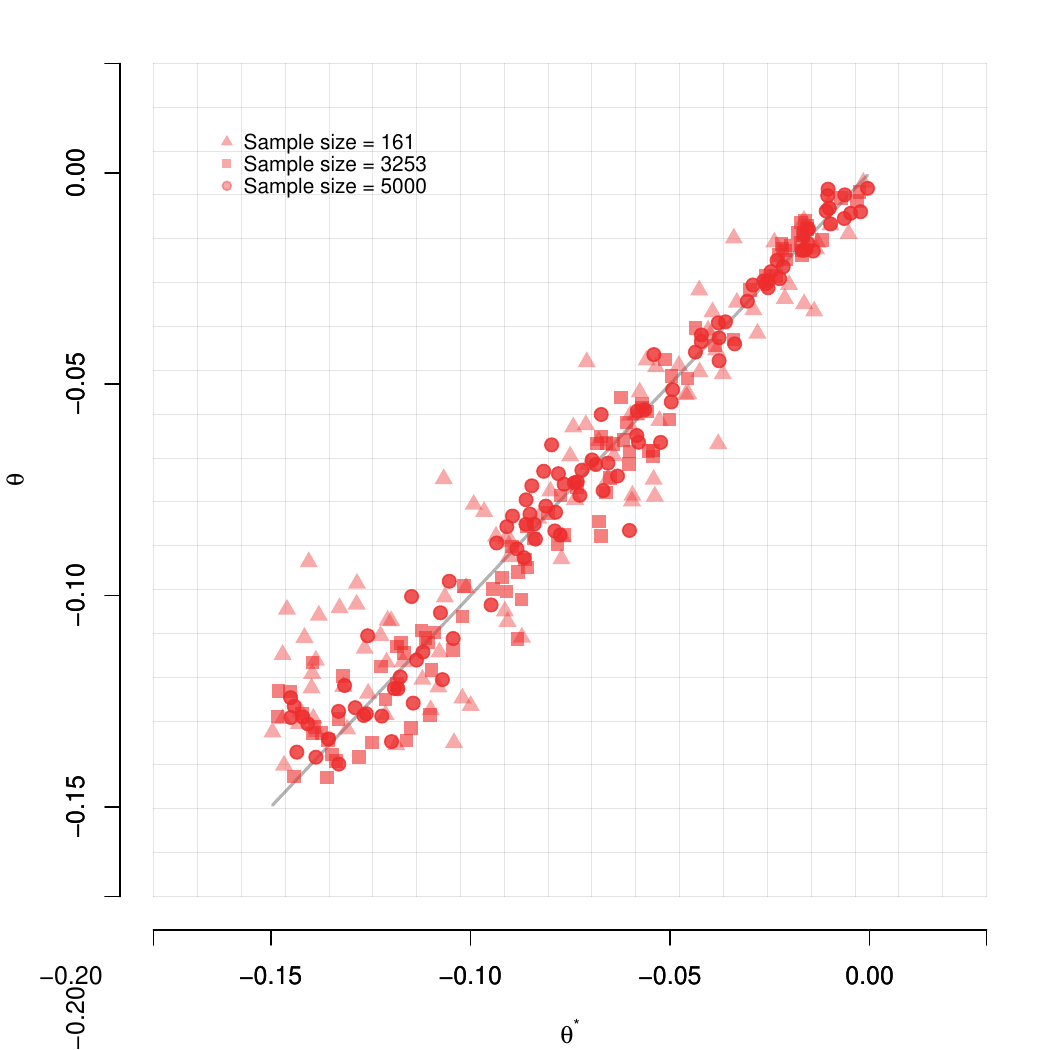}
                \caption{$\phi_{2}$}
        \end{subfigure}
        \begin{subfigure}[t]{0.33\textwidth}
                \centering
               \includegraphics[width=\textwidth]{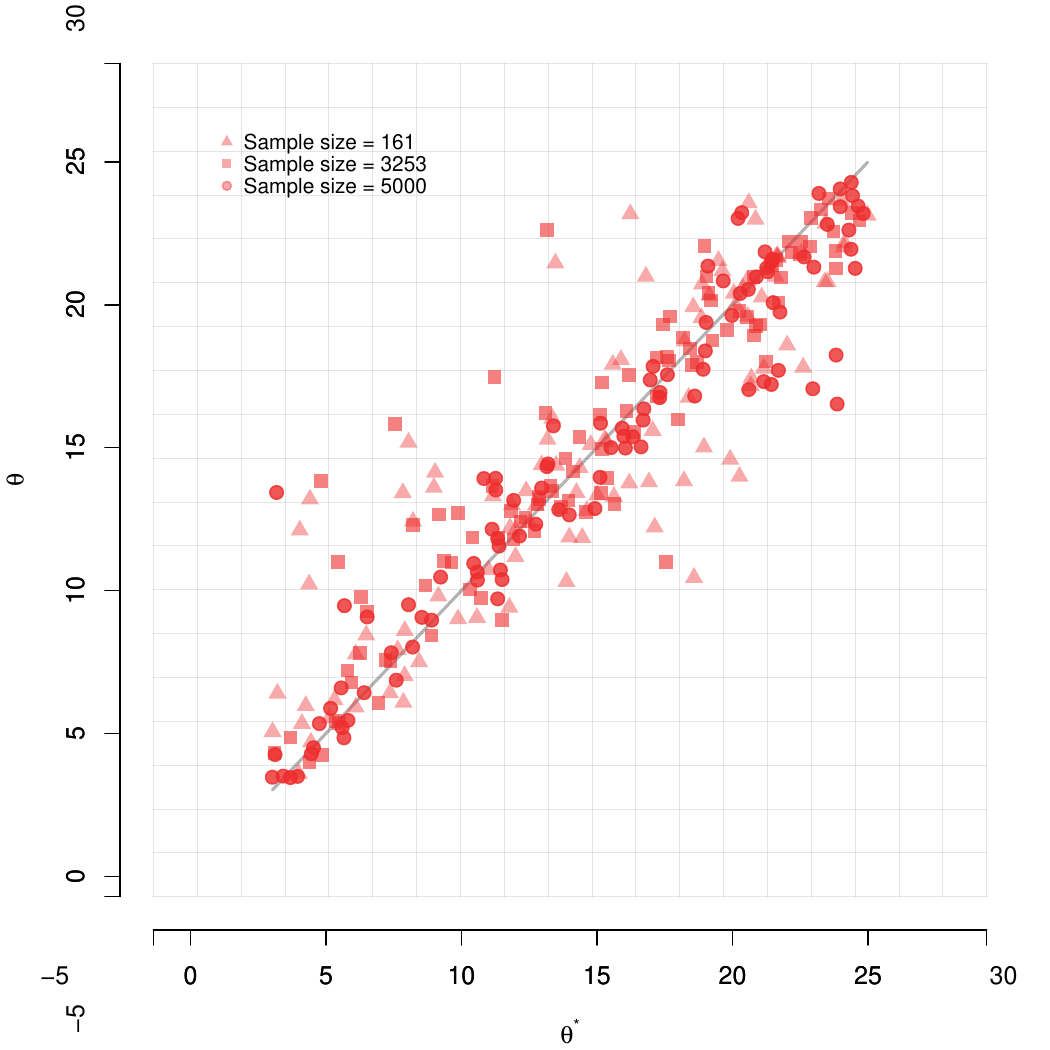}
                \caption{$\delta$}
        \end{subfigure}
  \begin{subfigure}[t]{0.33\textwidth}
                \centering
               \includegraphics[width=\textwidth]{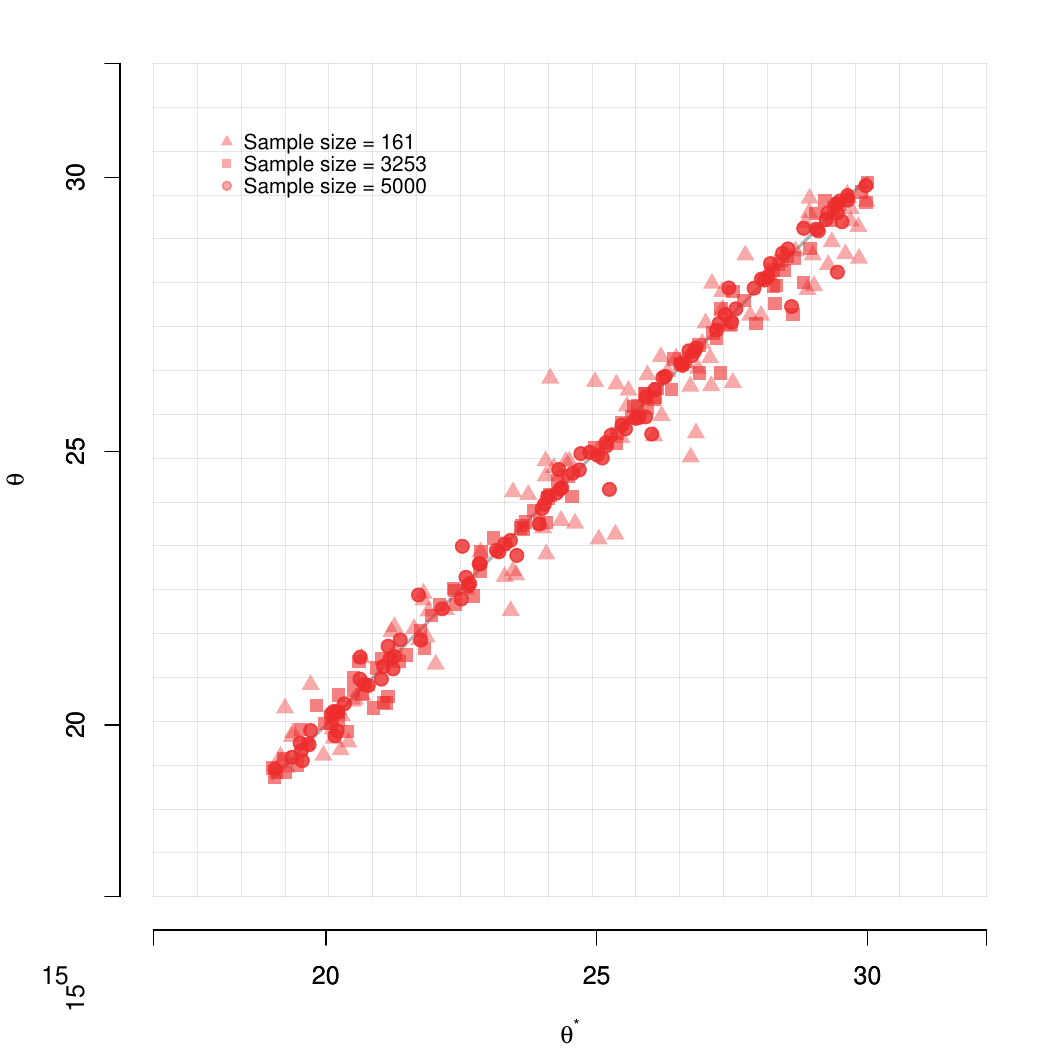}
                \caption{$\mu_m$}
        \end{subfigure}
          \begin{subfigure}[t]{0.33\textwidth}
                \centering
               \includegraphics[width=\textwidth]{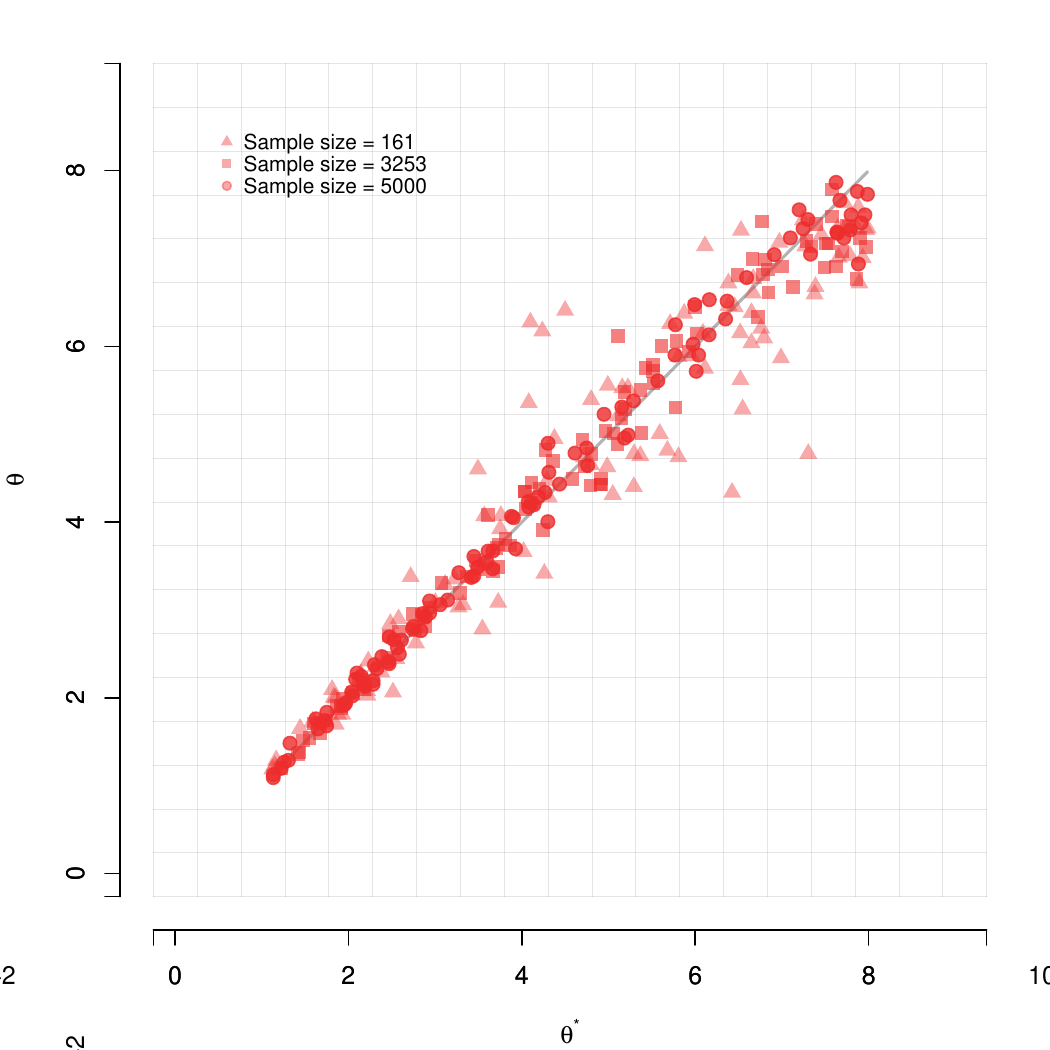}
                \caption{$\sigma_{m}$}
        \end{subfigure}
	\label{cval}
\end{figure}

The results of the cross-validation exercise are shown in Figure~\ref{cval}. The three sets of results correspond to the different sample sizes used in the simulations. The results obtained with the largest sample size show that the estimation procedure consistently returns parameter values $\theta$ that are very close to the true parameters that generated the data $\theta^*$. As expected, simulations results obtained with a smaller number of marriages are associated with large preduction errors, which shows the estimation approach correctly reflects sample size uncertainty. 

The parameter with the smallest associated prediction errors is $\mu_m$ and the parameter with the largest associated errors is $\delta$. This is not unexpected as a change in the value of $\mu_{m}$ produces a shift of the entire distribution of rates that is very distinct from the effect of a value change in any other parameter. A change in the value of $\delta$, on the other hand, produces an effect on the rates that overlaps to a certain extent to the effect of a change in the value of $\phi_{1}$ (see appendix~\ref{ap.A} for some animations illustrating the effect of each parameters on the schedule of age-specific fertility rates).

\floatsetup[figure]{capposition=top}

\subsection{Model Fit}

\begin{figure} [H]
\caption[Observed vs. Simulated Age-Specific Fertility Rates with 95\% Credible Intervals.]{\textbf{Observed vs. Simulated Age-Specific Fertility Rates with 95\% Credible Intervals.}}
        \begin{subfigure}[t]{0.45\textwidth}
                \centering
                \includegraphics[width=\textwidth]{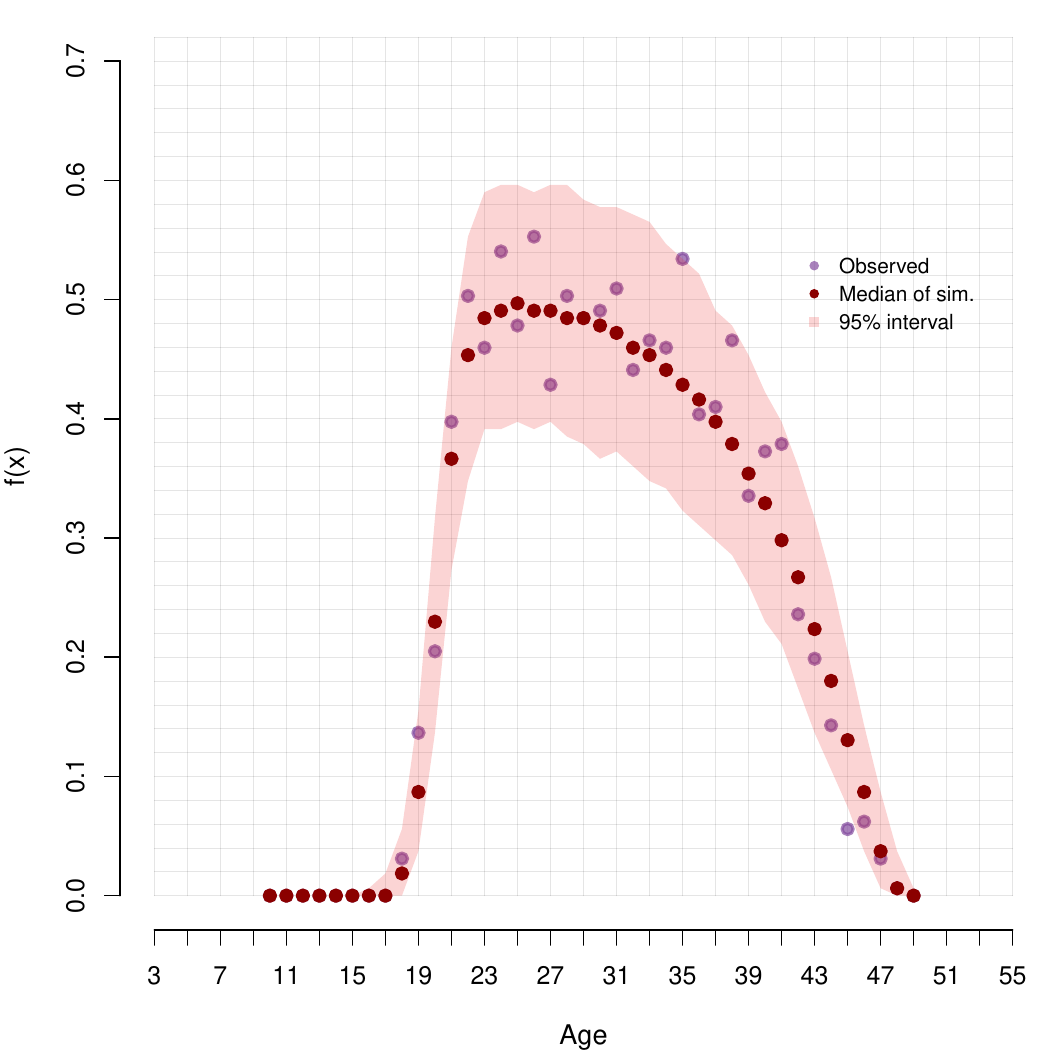}
                \caption{Hutterites}
                \label{fx_ht}
        \end{subfigure}%
        \begin{subfigure}[t]{0.45\textwidth}
                \centering
                \includegraphics[width=\textwidth]{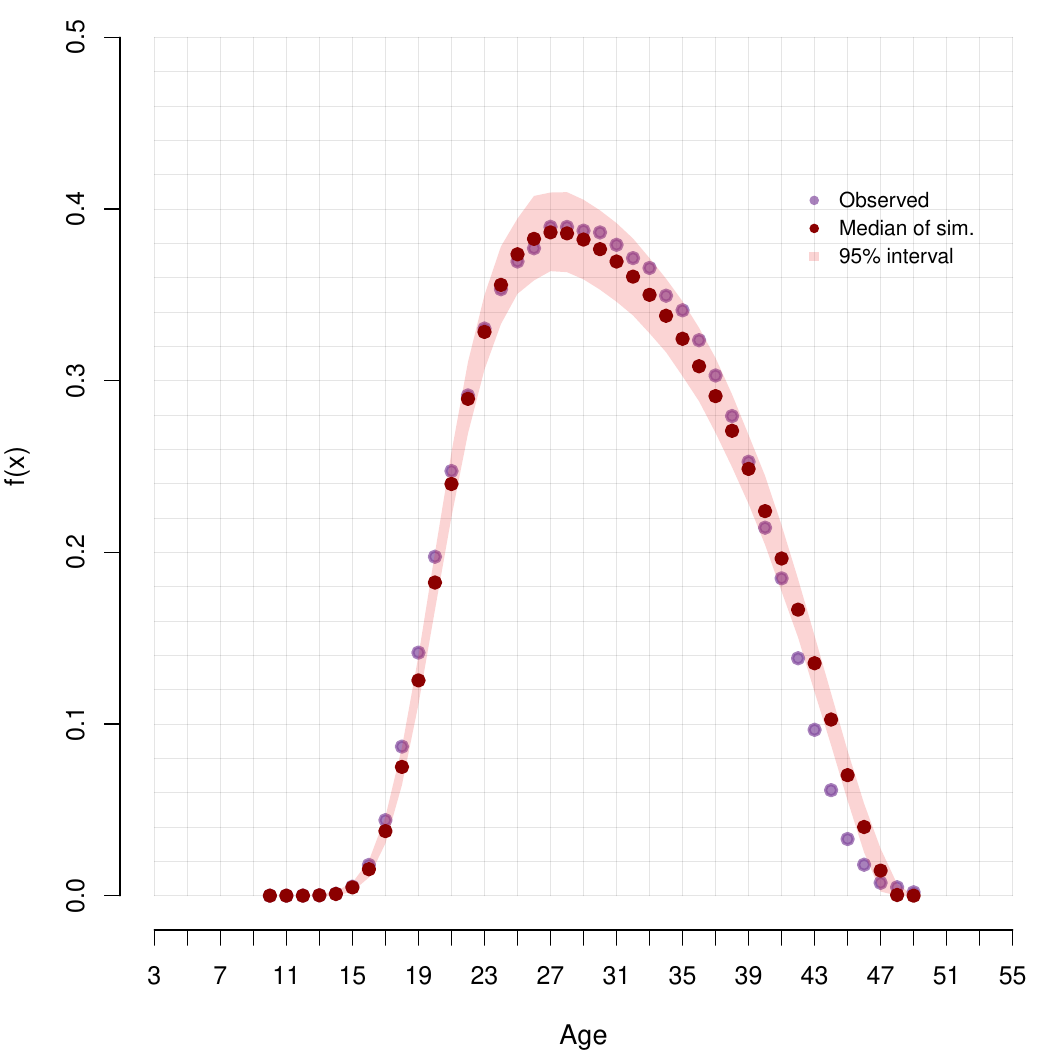}
                \caption{XVIII Century Quebec}
                \label{fx_FC}
        \end{subfigure}~

        \begin{subfigure}[t]{0.45\textwidth}
                \centering
               \includegraphics[width=\textwidth]{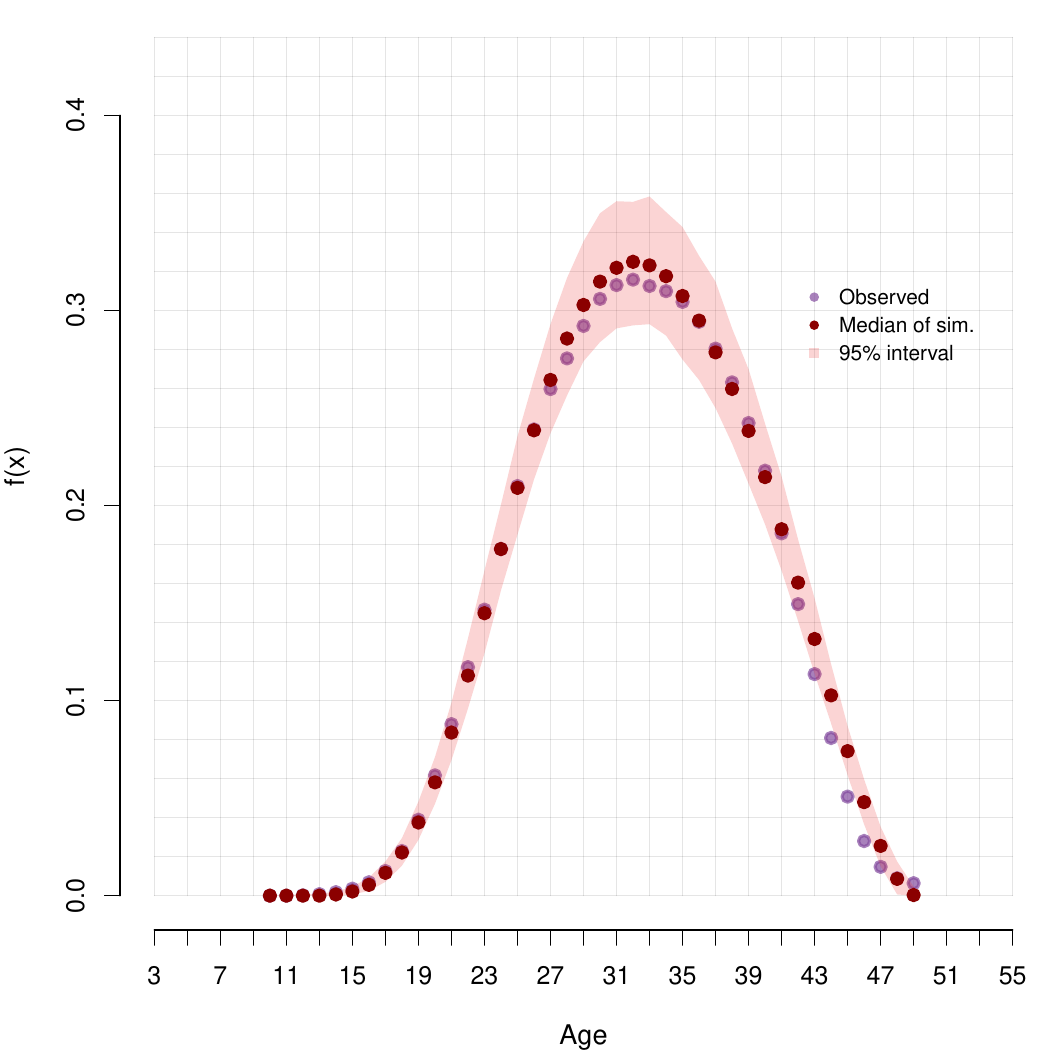}
                \caption{XVII - XVIII Century France}
                \label{fx_HF}
        \end{subfigure}

	\label{fit_fx}
\end{figure}

Figure~\ref{fit_fx} depicts the model's posterior predictive distribution against observed ASFR for the three populations analyzed. The model accurately captures the general characteristics of the data, such as the shape, location of peaks, and rate of decline of the three distributions. In fact, the fit is very satisfactory at most ages, except for the rates at the very end of the reproductive age window (ages 47 to 50), which the model tends to slightly overestimate. Credible intervals reflect the higher degree of uncertainty with which observed rates are estimated in the Hutterite's case.  


\floatsetup[figure]{capposition=top}

\section{Discussion}

The need to represent the age-distribution of fertility rates has overwhelmingly been met with solutions at the aggregate level. These solutions traded parameter interpretability in exchange for statistical formalization. Thanks to the approach we introduced in this article, this compromise is no longer necessary.

Our individual-level model is certainly a simple, even crude, description of the reproductive process. Yet, it fits a wide range of fertility schedules just as well as previous purely statistical models, with the major strength that it explicitly models the behavioral and biological mechanism that give rise to these schedules. 

Our approach also allows for the transparent incorporation of additional information via prior distributions when such information is available. This provides a very favorable contrast to the trail-and-error calibration procedures commonly used in micro-level models.

We believe these ideas have the potential to make contributions in at least two fields. In biodemography, e.g. in the study of non-human primates, our approach could be used to classify species in a meaningful way, by age at onset of sexual reproduction, postpartum amenorrhea duration and fecundability, rather than abstract properties of asfr curves. 

In demography, we hope it inspires the exploration of more complex models with the aim of  understanding the microfoundations of present and future fertility trends.

\appendix
\section{Appendix}
\label{ap.A}

\begin{figure}[H]
    \centering
    \caption{Effect of $\phi_{1}$ on Simulated Age-Specific Fertility Rates}
	\animategraphics[loop, controls=true,width=10cm]{6}{plots/phibeta1/plotseq}{1}{80}		
\end{figure}

\begin{figure}[H]
    \centering
 \caption{Effect of $\phi_{2}$ on Simulated Age-Specific Fertility Rates}
	\animategraphics[loop, controls=true,width=10cm]{6}{plots/phibeta2/plotseq}{1}{80}	
\end{figure}

\begin{figure}[H]
    \centering
 \caption{Effect of $\mu_m$ on Simulated Age-Specific Fertility Rates}
	\animategraphics[loop,controls=true,width=10cm]{6}{plots/mum/plotseq}{1}{80}	
\end{figure}

\begin{figure}[H]
    \centering	
 \caption{Effect of $\sigma_m$ on Simulated Age-Specific Fertility Rates}
	\animategraphics[loop,controls=true,width=10cm]{6}{plots/sigmam/plotseq}{1}{80}	
\end{figure}

\begin{figure}[H]
    \centering	
 \caption{Effect of $\delta$ on Simulated Age-Specific Fertility Rates}
	\animategraphics[loop, controls=true,width=10cm]{6}{plots/nsp/plotseq}{1}{80}	
\end{figure}

\section{Appendix}
\label{ap.B}

\floatsetup[figure]{capposition=bottom}
\begin{figure}[H]
  \setlength\tabcolsep{6pt}
  \adjustboxset{width=\linewidth,valign=c}
  \centering
  \begin{tabularx}{1.0\linewidth}{@{}
      l
      X @{\hspace{6pt}}
      X
      X
    @{}}
    & \multicolumn{1}{c}{\textbf{$\phi_{1}$}}
    & \multicolumn{1}{c}{\textbf{$\phi_{2}$}}
    & \multicolumn{1}{c}{\textbf{$\delta$}}  \\
    \rotatebox[origin=c]{90}{\textbf{Hutterites}}
    & \includegraphics{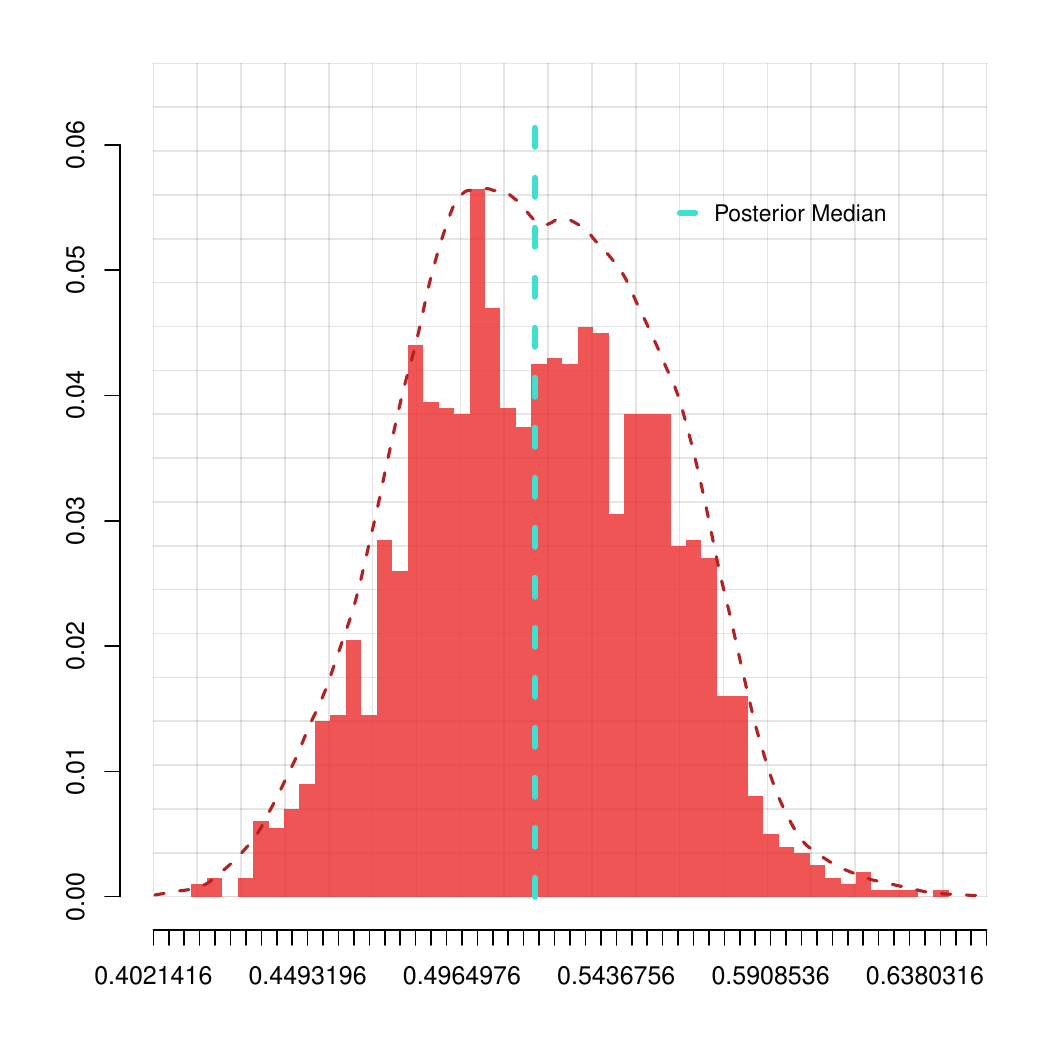}
    & \includegraphics{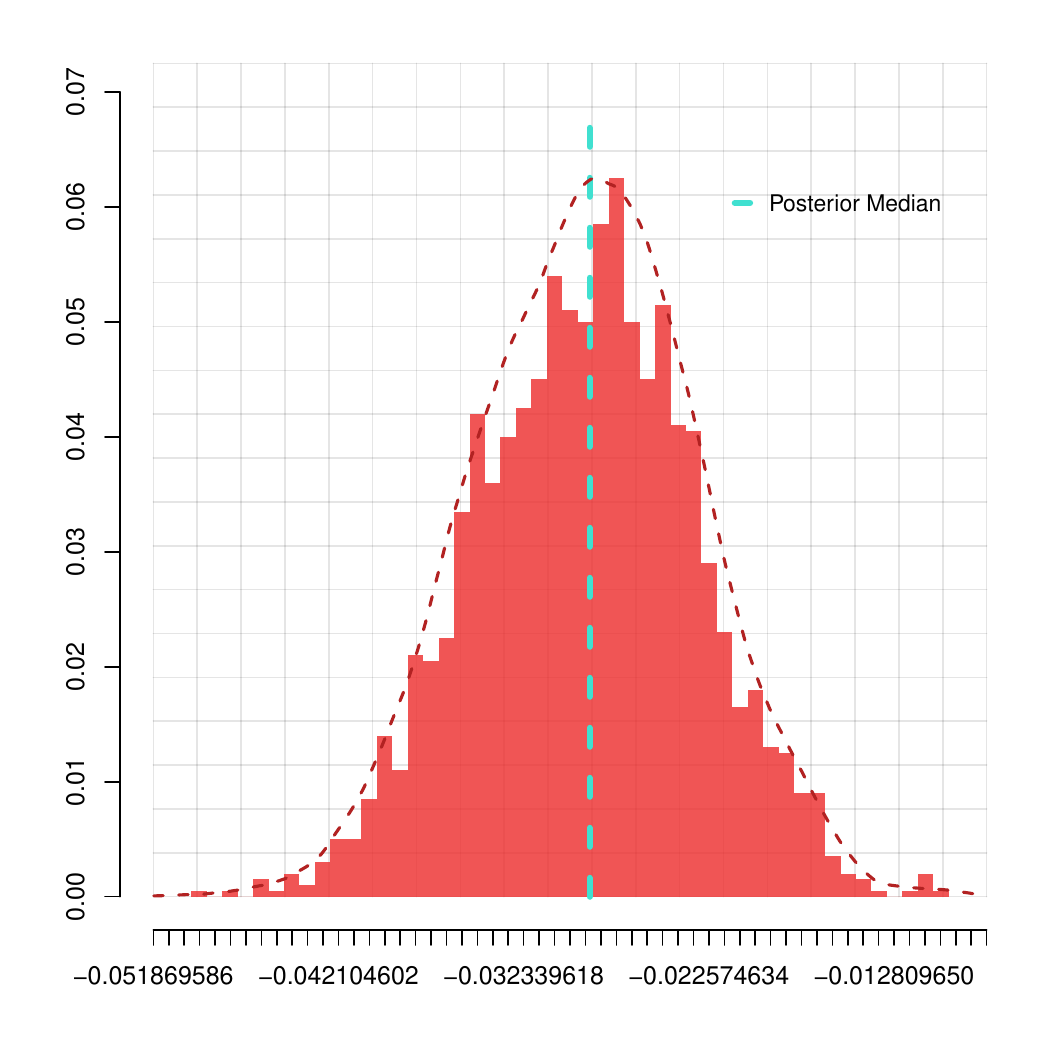}
    & \includegraphics{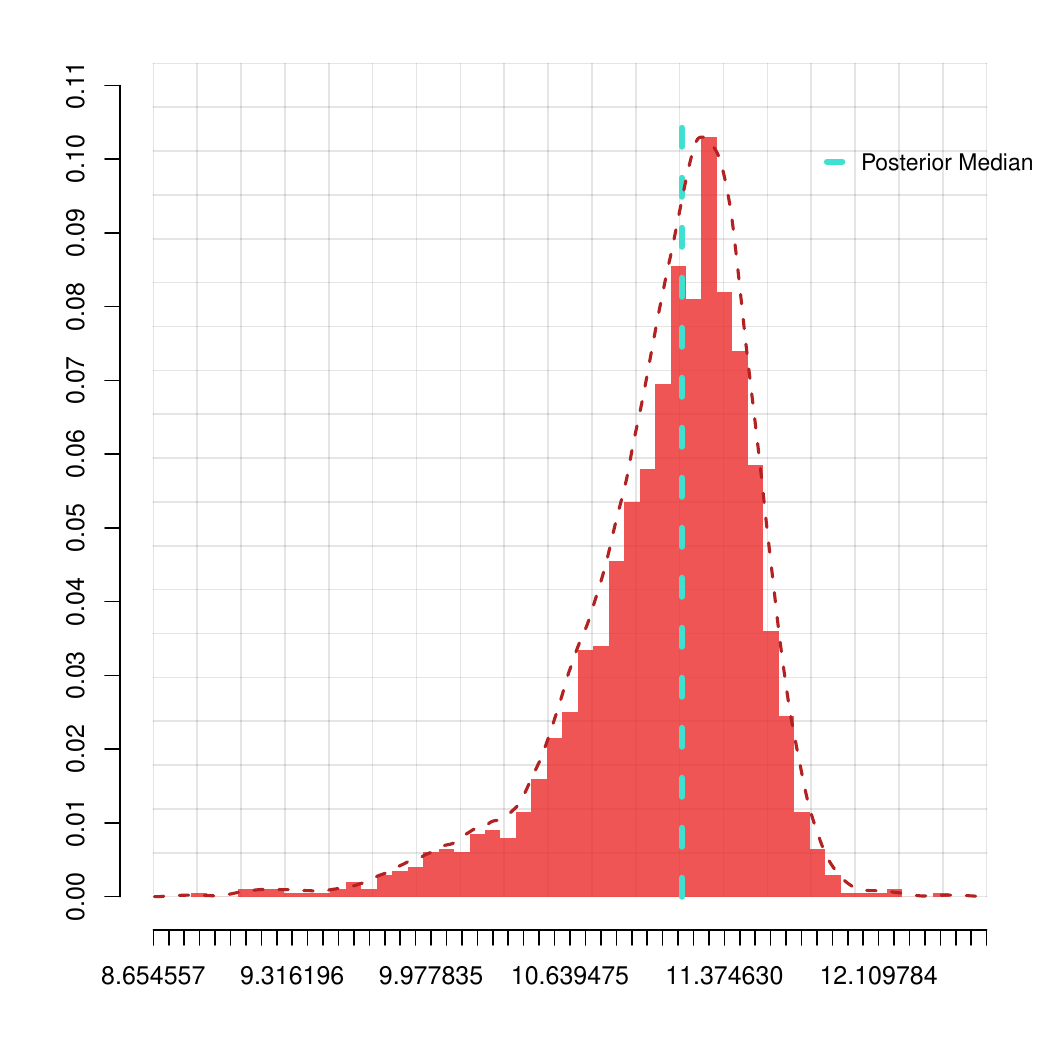}  \\
    \rotatebox[origin=c]{90}{\textbf{French Canadian}}
    & \includegraphics{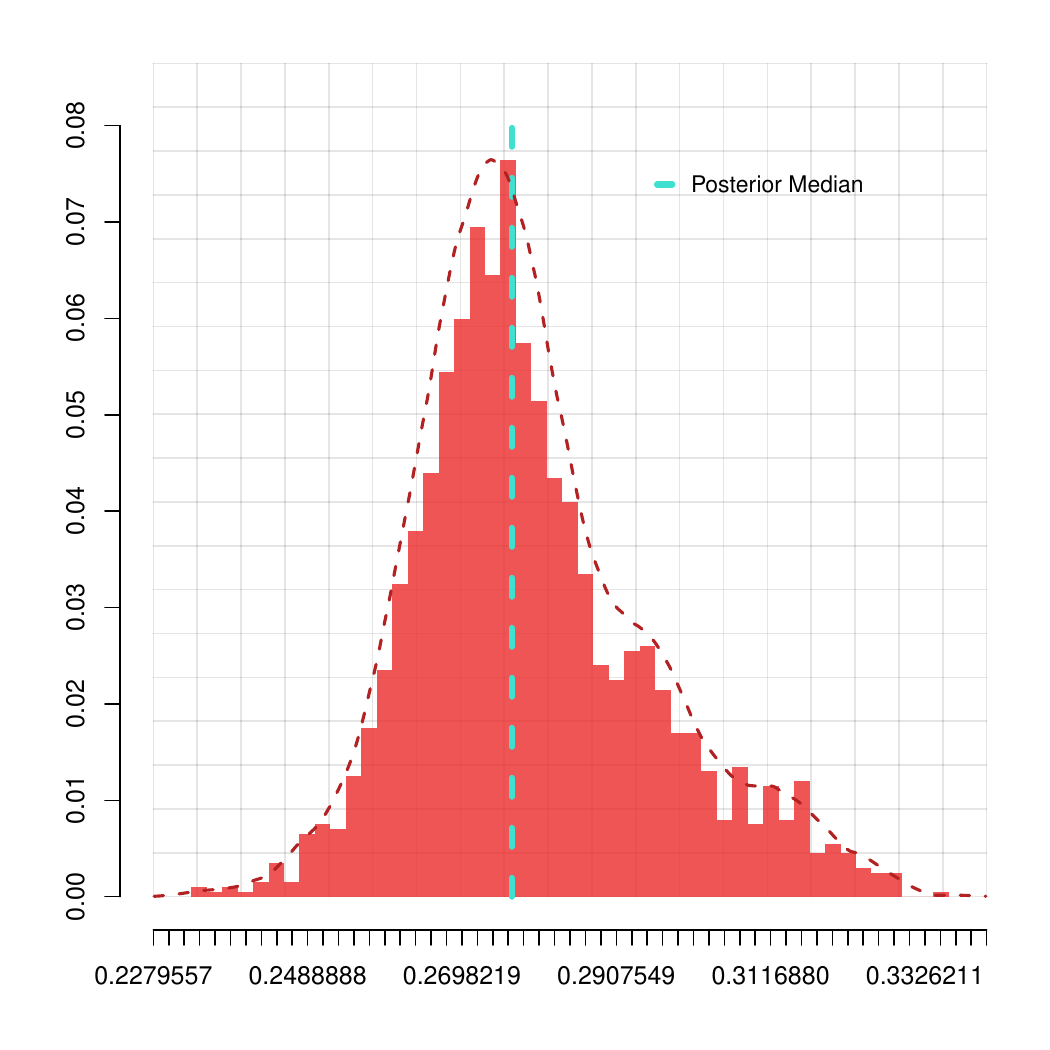}
    & \includegraphics{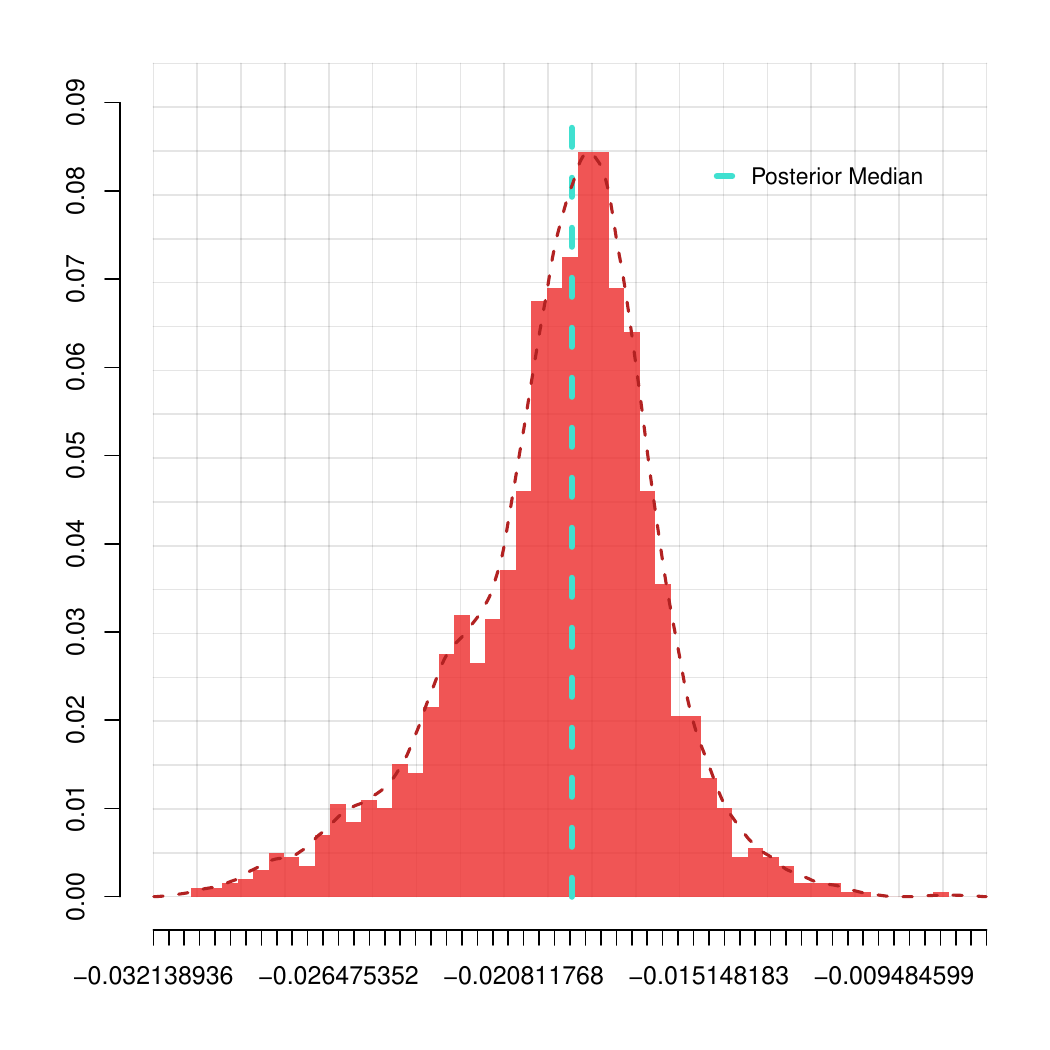}
    & \includegraphics{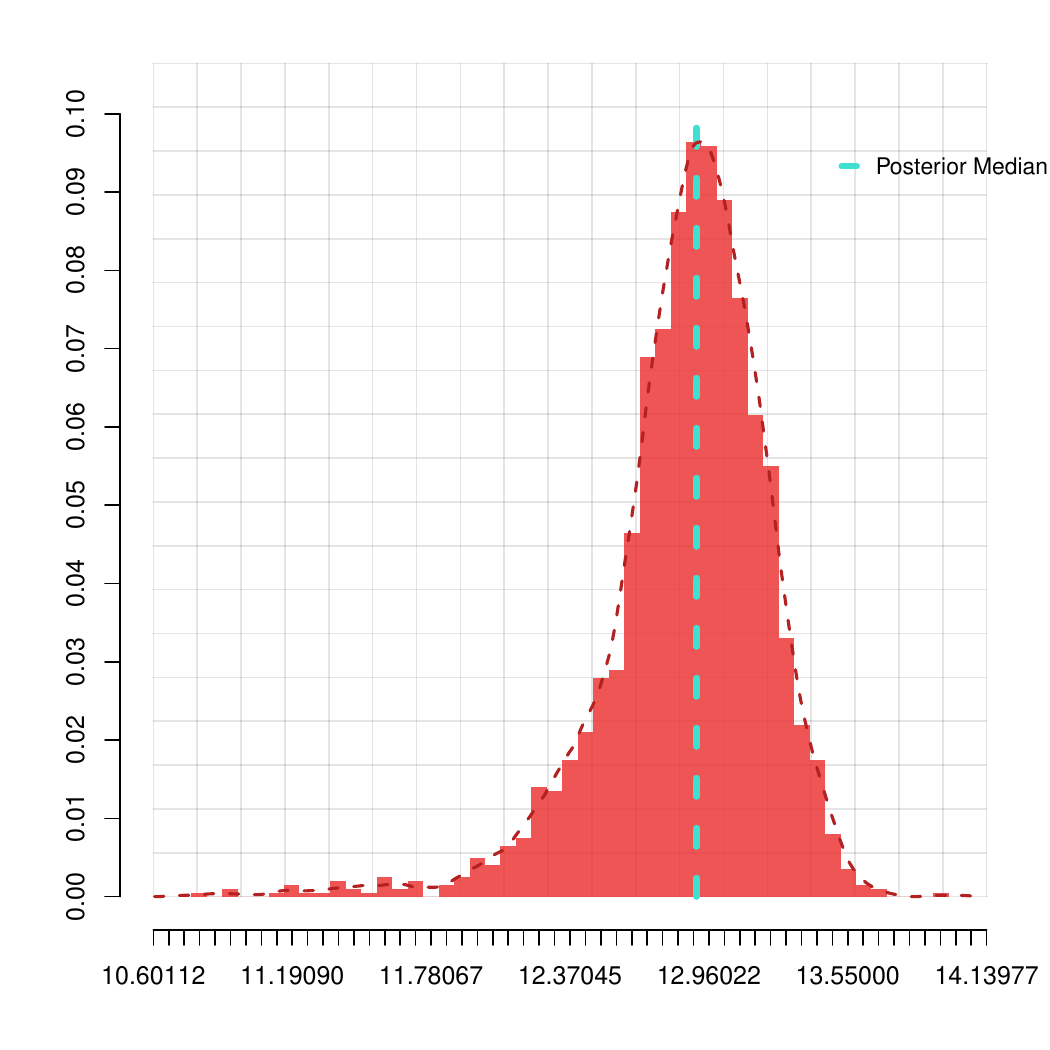} \\
    \rotatebox[origin=c]{90}{\textbf{Historic France}}
    & \includegraphics{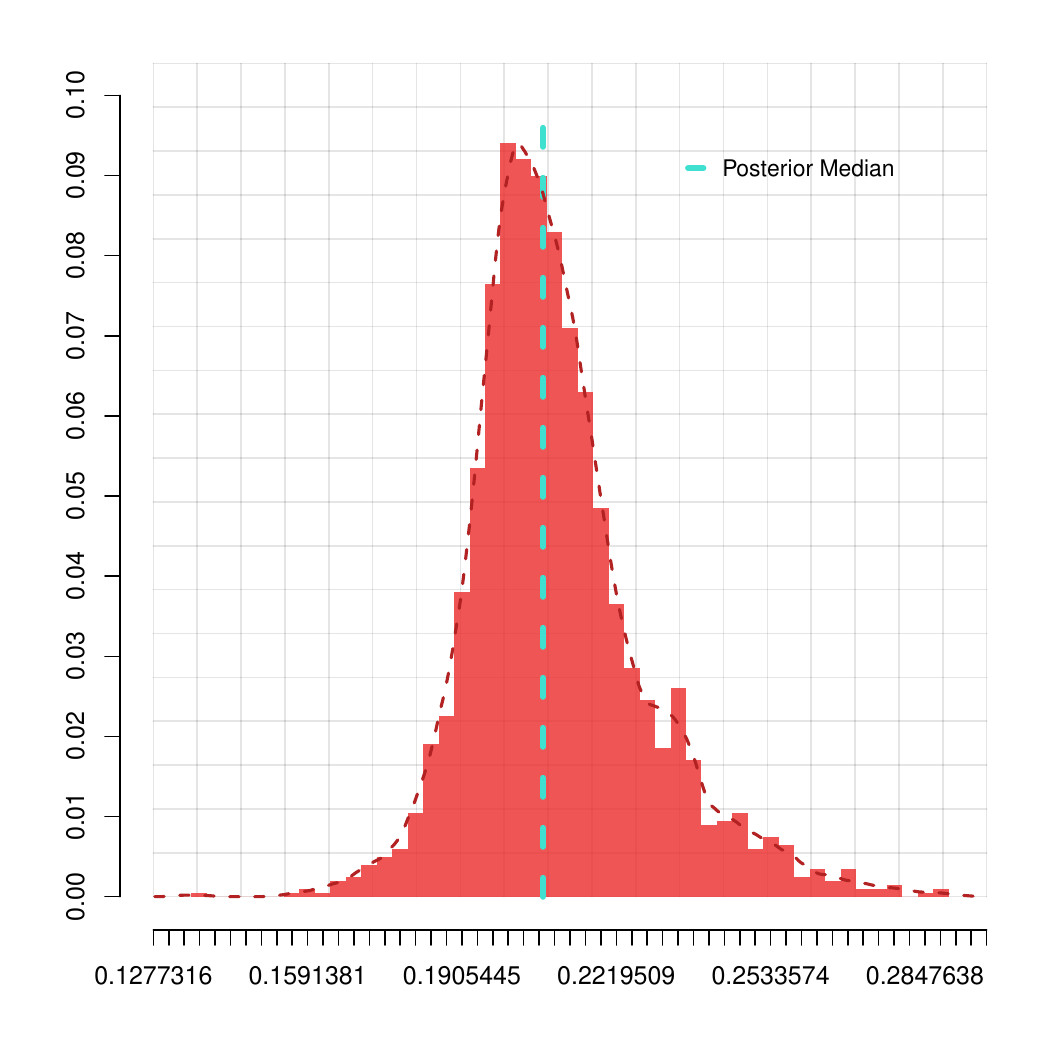}
    & \includegraphics{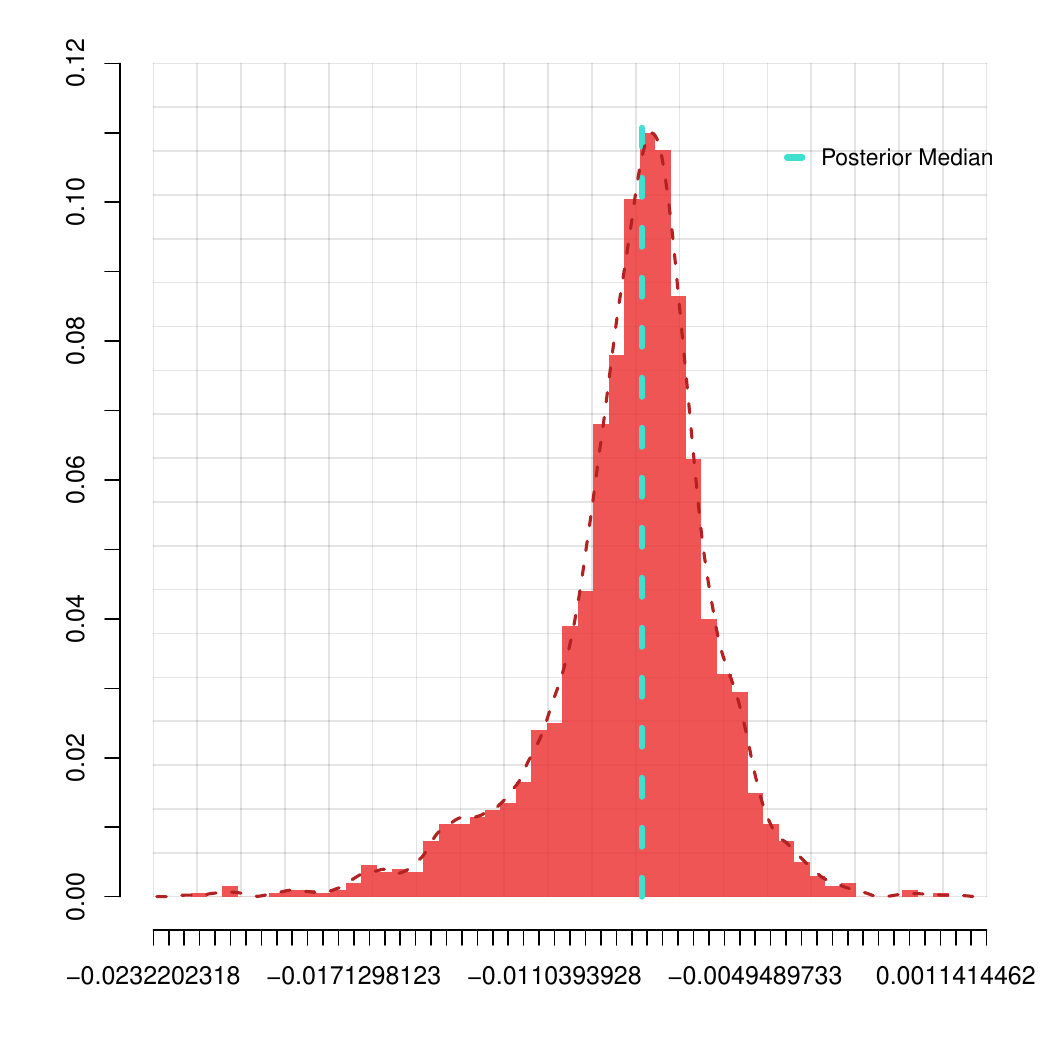}
    & \includegraphics{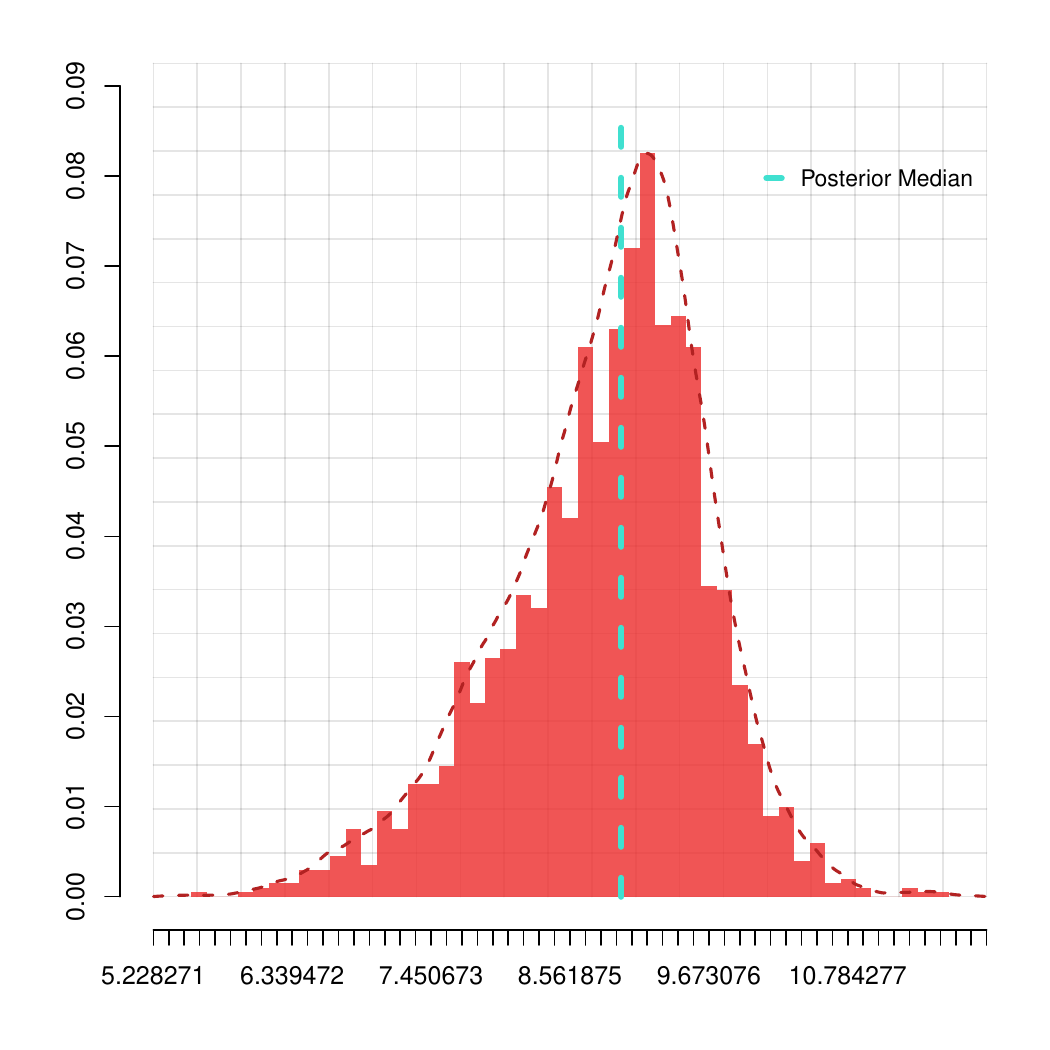}  \\
  \end{tabularx}
  \caption{Adjusted Posterior Distributions}
  	\label{post_nat}
\end{figure}

\floatsetup[figure]{capposition=bottom}
\begin{figure}[H]
  \setlength\tabcolsep{6pt}
  \adjustboxset{width=\linewidth,valign=c}
  \centering
  \begin{tabularx}{1.0\linewidth}{@{}
      l
      X @{\hspace{6pt}}
      X
    @{}}
    & \multicolumn{1}{c}{\textbf{$\mu_{m}$}} 
    & \multicolumn{1}{c}{\textbf{$\sigma_{m}$}}  \\
    \rotatebox[origin=c]{90}{\textbf{Hutterites}}
    & \includegraphics{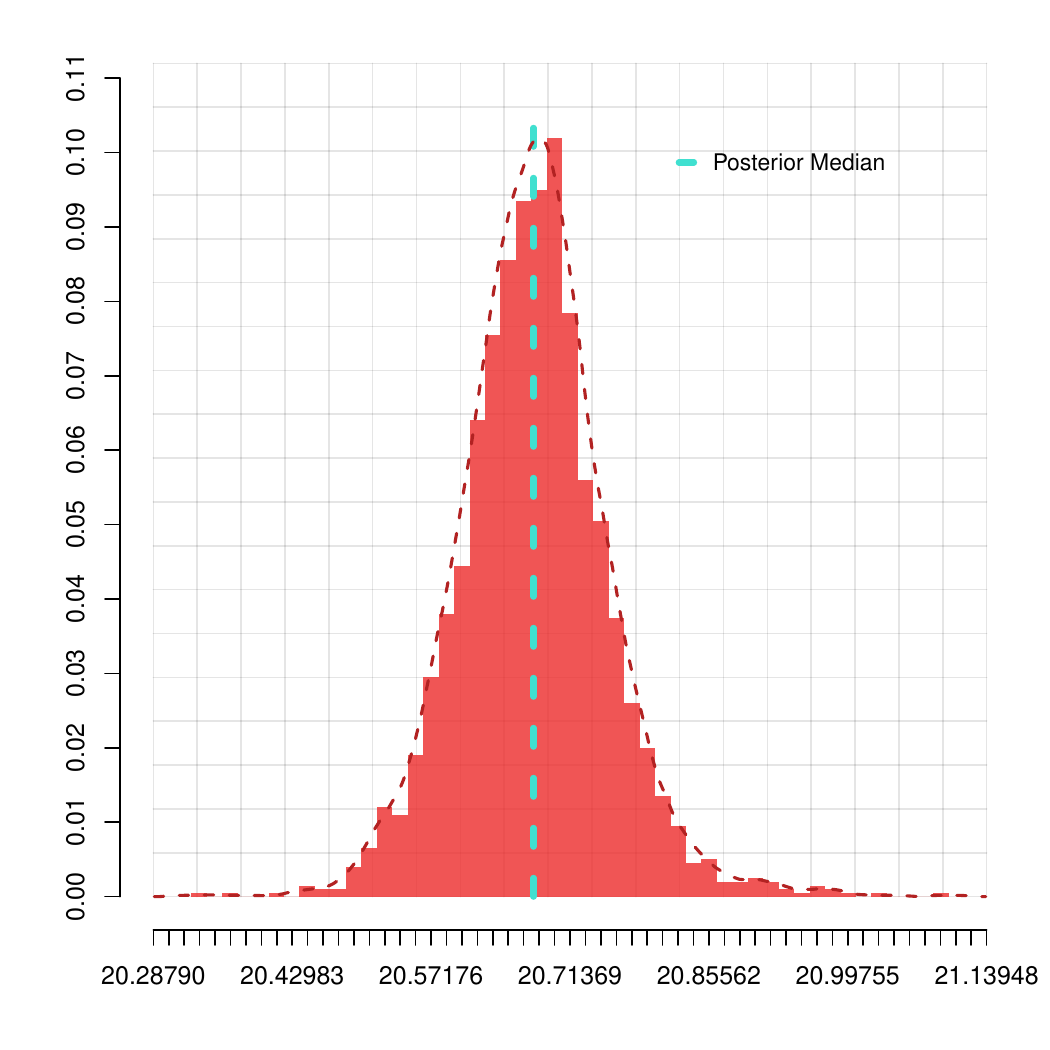} 
    & \includegraphics{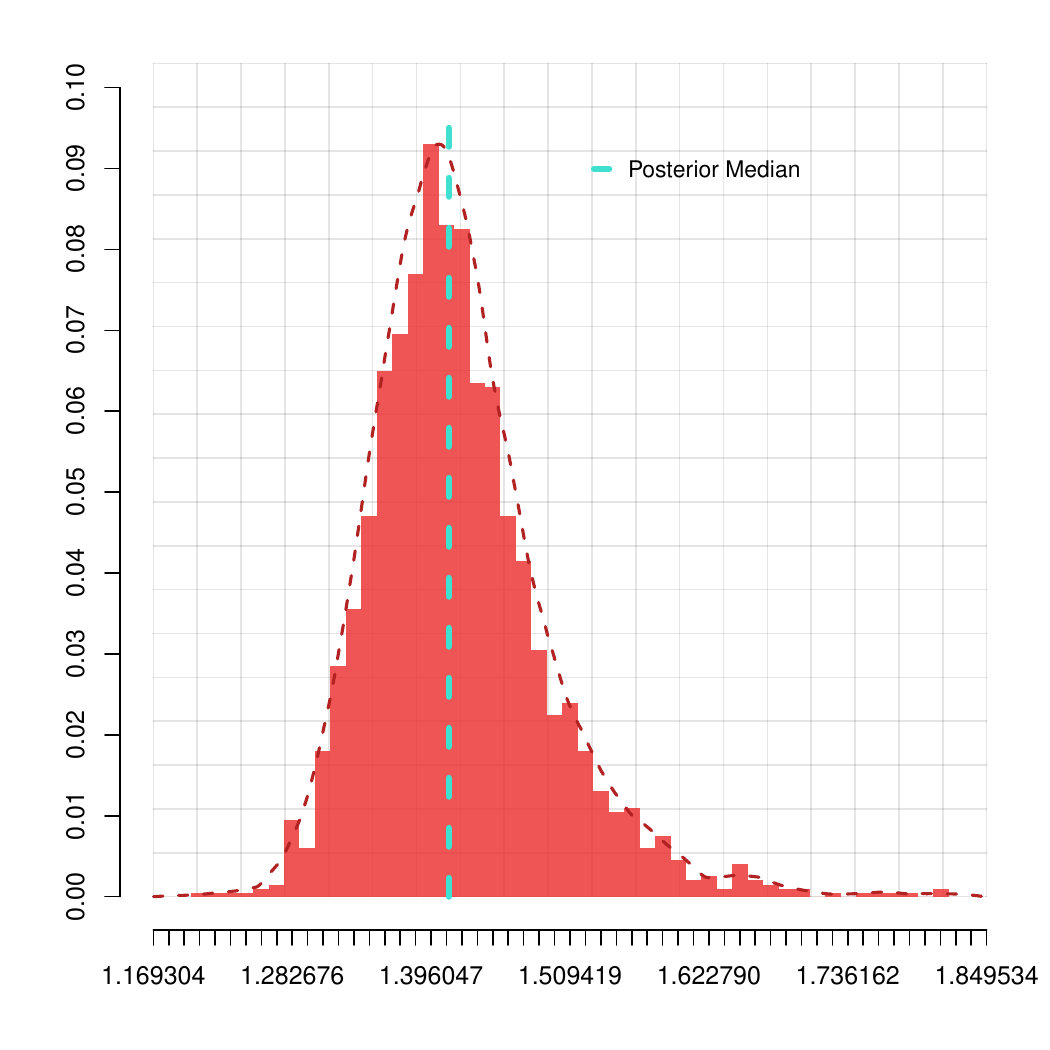}  \\
    \rotatebox[origin=c]{90}{\textbf{French Canadian}}
    & \includegraphics{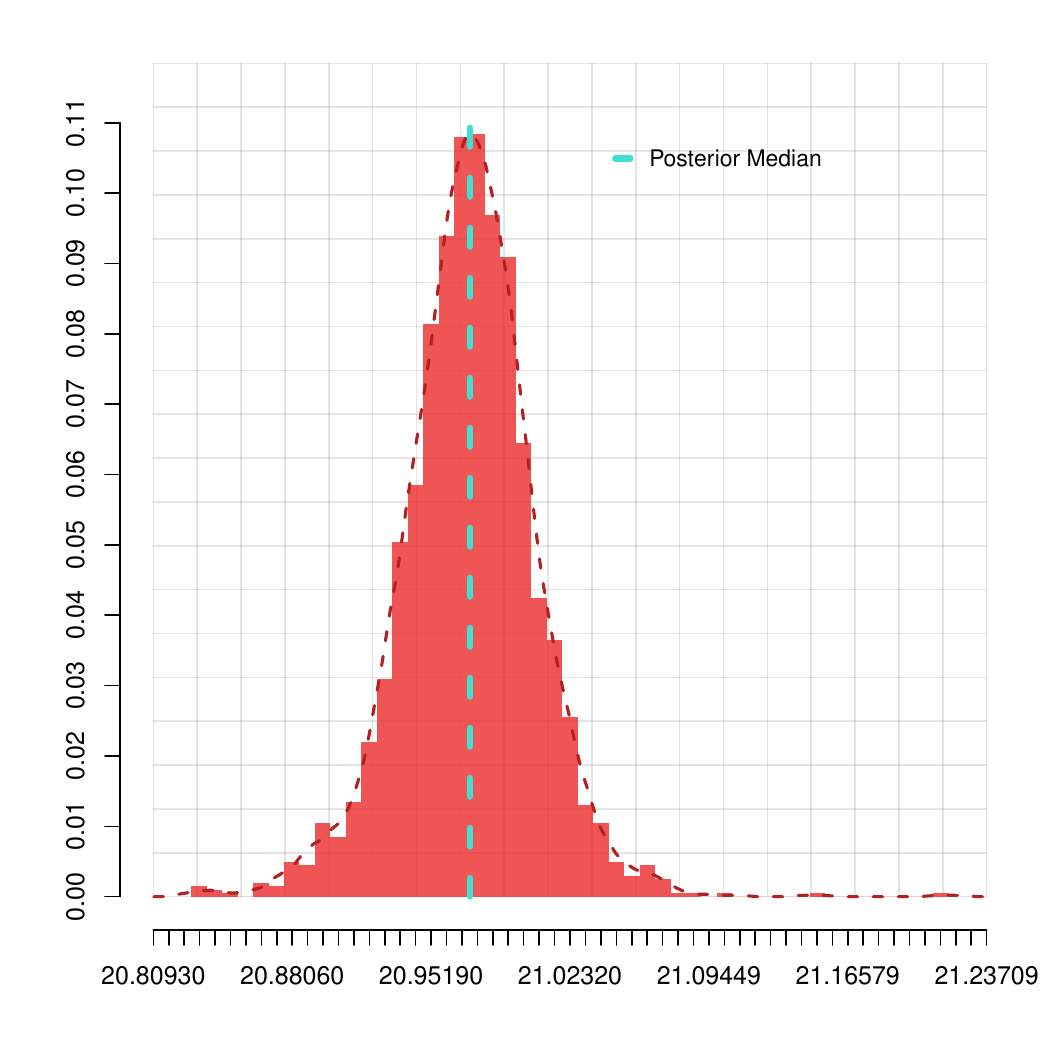} 
    & \includegraphics{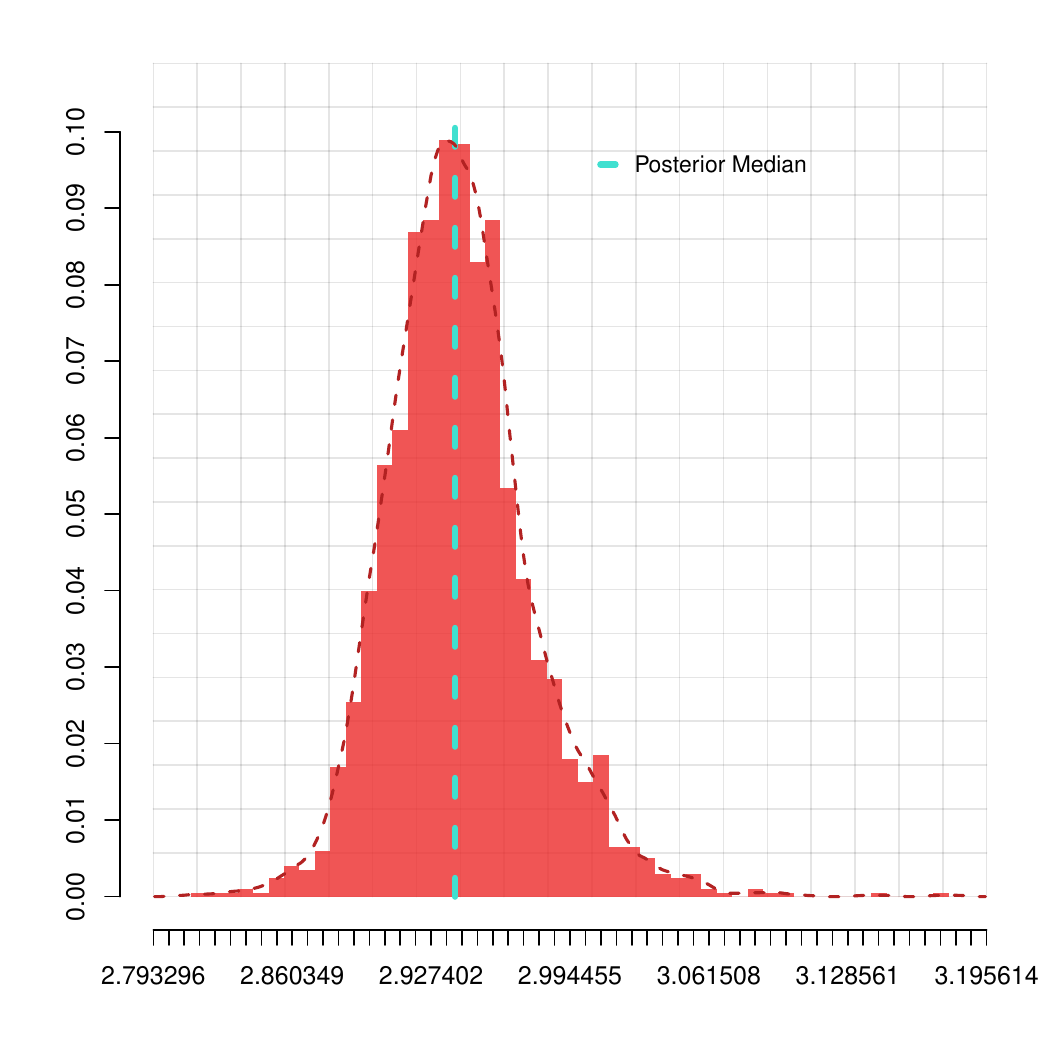}  \\
    \rotatebox[origin=c]{90}{\textbf{Historic France}}
    & \includegraphics{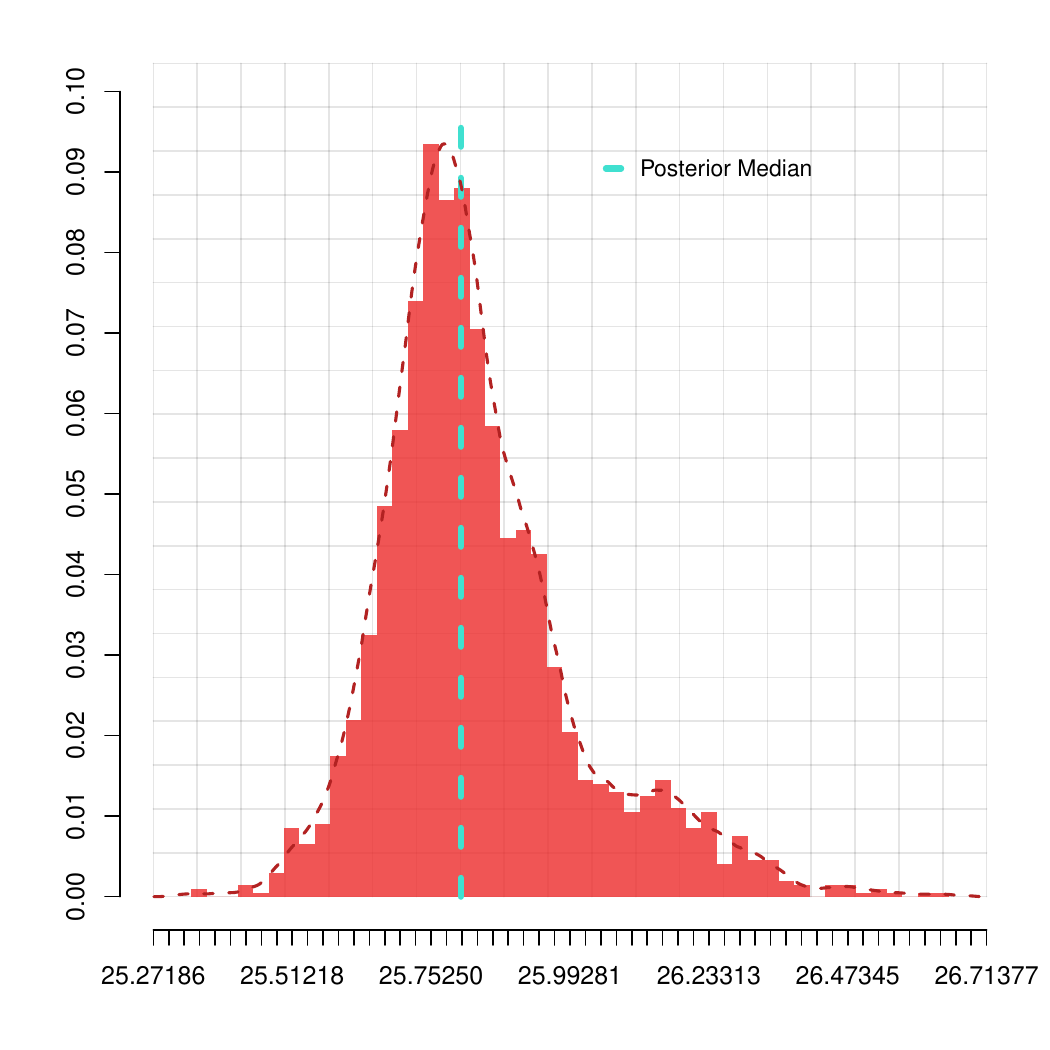} 
    & \includegraphics{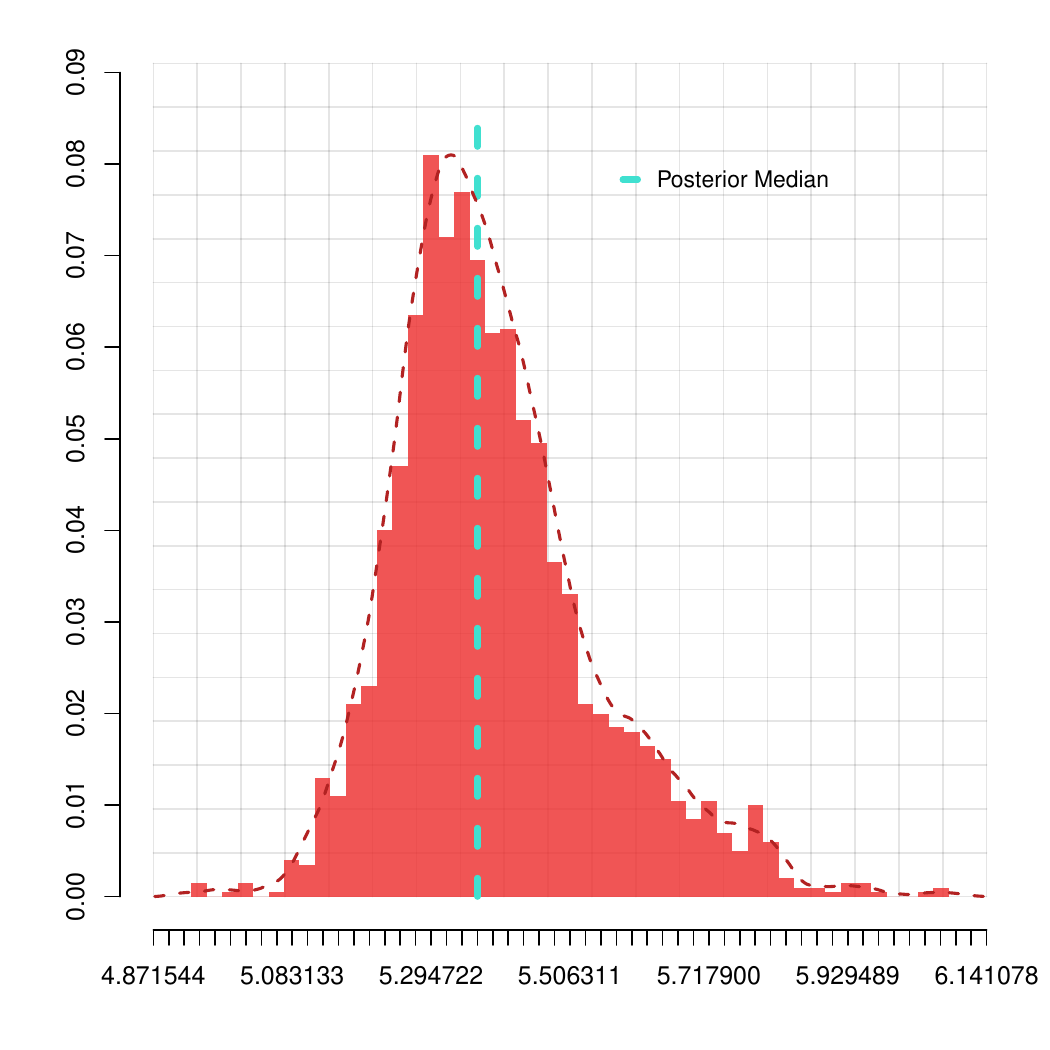}  \\
  \end{tabularx}
  \caption{Adjusted Posterior Distributions}
  	\label{post_nat}
\end{figure}

\bibliographystyle{chicago}
\bibliography{biblio}

\newcommand{\noop}[1]{}
\begin{thebibliography}{}

\bibitem[\protect\citeauthoryear{Barrett}{Barrett}{1971}]{Barrett1971}
Barrett, J.~C. (1971).
\newblock Use of a fertility simulation model to refine measurement techniques.
\newblock {\em Demography\/}~{\em 8\/}(4), 481--490.

\bibitem[\protect\citeauthoryear{Beaumont}{Beaumont}{2019}]{Beaumont2019}
Beaumont, M.~A. (2019).
\newblock Approximate bayesian computation.
\newblock {\em Annual review of statistics and its application\/}~{\em 6},
  379--403.

\bibitem[\protect\citeauthoryear{Blum}{Blum}{2018}]{blum2018regression}
Blum, M.~G. (2018).
\newblock Regression approaches for abc.
\newblock {\em Handbook of approximate Bayesian computation\/}, 71--85.

\bibitem[\protect\citeauthoryear{Clark, Cummins, and Curtis}{Clark
  et~al.}{2020}]{Clark2020}
Clark, G., N.~Cummins, and M.~Curtis (2020).
\newblock Twins support the absence of parity-dependent fertility control in
  pretransition populations.
\newblock {\em Demography\/}~{\em 57\/}(4), 1571--1595.

\bibitem[\protect\citeauthoryear{Eaton and Mayer}{Eaton and
  Mayer}{1953}]{Eaton1953}
Eaton, J.~W. and A.~J. Mayer (1953).
\newblock The social biology of very high fertility among the hutterites. the
  demography of a unique population.
\newblock {\em Human biology\/}~{\em 25\/}(3), 206.

\bibitem[\protect\citeauthoryear{Eijkemans, Van~Poppel, Habbema, Smith,
  Leridon, and Te~Velde}{Eijkemans et~al.}{2014}]{Eijkemans2014}
Eijkemans, M.~J., F.~Van~Poppel, D.~F. Habbema, K.~R. Smith, H.~Leridon, and
  E.~R. Te~Velde (2014).
\newblock Too old to have children? lessons from natural fertility populations.
\newblock {\em Human Reproduction\/}~{\em 29\/}(6), 1304--1312.

\bibitem[\protect\citeauthoryear{Gini}{Gini}{1924}]{gini1924premieres}
Gini, C. (1924).
\newblock Premi{\`e}res recherches sur la f{\'e}condabilit{\'e} de la femme.
\newblock In North-Holland: (Ed.), {\em Proceedings of the International
  Mathematical Congress.}, Volume Vol. 2., Toronto.

\bibitem[\protect\citeauthoryear{Henry}{Henry}{1953}]{Henry1953}
Henry, L. (1953).
\newblock Fondements théoriques des mesures de la fécondité naturelle.
\newblock {\em {Revue de l'Institut International de Statistique / Review of
  the International Statistical Institute}\/}~{\em 21\/}(3), 135--151.

\bibitem[\protect\citeauthoryear{Hoem, Madsen, Nielsen, Ohlsen, Hansen, and
  Rennermalm}{Hoem et~al.}{1981}]{hoem1981experiments}
Hoem, J.~M., D.~Madsen, J.~L. Nielsen, E.-M. Ohlsen, H.~O. Hansen, and
  B.~Rennermalm (1981).
\newblock Experiments in modelling recent danish fertility curves.
\newblock {\em Demography\/}~{\em 18\/}(2), 231--244.

\bibitem[\protect\citeauthoryear{Ingoldsby and Stanton}{Ingoldsby and
  Stanton}{1988}]{Ingoldsby1988}
Ingoldsby, B.~B. and M.~E. Stanton (1988).
\newblock The hutterites and fertility control.
\newblock {\em Journal of Comparative Family Studies\/}~{\em 19\/}(1),
  137--142.

\bibitem[\protect\citeauthoryear{Larsen and Yan}{Larsen and
  Yan}{2000}]{Larsen2000}
Larsen, U. and S.~Yan (2000).
\newblock The age pattern of fecundability: an analysis of french canadian and
  hutterite birth histories.
\newblock {\em Social biology\/}~{\em 47\/}(1-2), 34--50.

\bibitem[\protect\citeauthoryear{Le~Bras}{Le~Bras}{1993}]{le1993simulation}
Le~Bras, H. (1993).
\newblock {\em Simulation of change to validate demographic analysis.}
\newblock Oxford England Clarendon Press 1993.

\bibitem[\protect\citeauthoryear{Lee and Brattrud}{Lee and
  Brattrud}{1967}]{Lee1967}
Lee, S. and A.~Brattrud (1967).
\newblock Marriage under a monastic mode of life: A preliminary report on the
  hutterite family in south dakota.
\newblock {\em Journal of Marriage and the Family\/}, 512--520.

\bibitem[\protect\citeauthoryear{Mange}{Mange}{1964}]{Mange1964}
Mange, A.~P. (1964).
\newblock Growth and inbreeding of a human isolate.
\newblock {\em Human Biology\/}~{\em 36\/}(2), 104--133.

\bibitem[\protect\citeauthoryear{Pritchard, Seielstad, Perez-Lezaun, and
  Feldman}{Pritchard et~al.}{1999}]{pritchard1999population}
Pritchard, J.~K., M.~T. Seielstad, A.~Perez-Lezaun, and M.~W. Feldman (1999).
\newblock Population growth of human y chromosomes: a study of y chromosome
  microsatellites.
\newblock {\em Molecular biology and evolution\/}~{\em 16\/}(12), 1791--1798.

\bibitem[\protect\citeauthoryear{Ridley and Sheps}{Ridley and
  Sheps}{1966}]{ridley1966analytic}
Ridley, J.~C. and M.~C. Sheps (1966).
\newblock An analytic simulation model of human reproduction with demographic
  and biological components.
\newblock {\em Population Studies\/}~{\em 19\/}(3), 297--310.

\bibitem[\protect\citeauthoryear{Rubin}{Rubin}{1984}]{rubin1984bayesianly}
Rubin, D.~B. (1984).
\newblock Bayesianly justifiable and relevant frequency calculations for the
  applied statistician.
\newblock {\em The Annals of Statistics\/}, 1151--1172.

\bibitem[\protect\citeauthoryear{Sheps}{Sheps}{1965}]{Sheps1965}
Sheps, M.~C. (1965).
\newblock An analysis of reproductive patterns in an american isolate.
\newblock {\em Population studies\/}~{\em 19\/}(1), 65--80.

\bibitem[\protect\citeauthoryear{Sheps, Menken, and Radick}{Sheps
  et~al.}{1973}]{sheps1973mathematical}
Sheps, M.~C., J.~A. Menken, and A.~P. Radick (1973).
\newblock {\em Mathematical models of conception and birth}.
\newblock University of Chicago Press Chicago.

\bibitem[\protect\citeauthoryear{Tavar{\'e}, Balding, Griffiths, and
  Donnelly}{Tavar{\'e} et~al.}{1997}]{tavare1997inferring}
Tavar{\'e}, S., D.~J. Balding, R.~C. Griffiths, and P.~Donnelly (1997).
\newblock Inferring coalescence times from dna sequence data.
\newblock {\em Genetics\/}~{\em 145\/}(2), 505--518.

\bibitem[\protect\citeauthoryear{V{\'e}zina and Bournival}{V{\'e}zina and
  Bournival}{2020}]{Vezina2020}
V{\'e}zina, H. and J.-S. Bournival (2020).
\newblock An overview of the balsac population database: past developments,
  current state and future prospects.
\newblock {\em Historical Life Course Studies\/}~{\em 9}, 114--129.

\end{thebibliography}

\end{document}